\documentclass[aps,prd,showpacs,nofootinbib,showkeys]{revtex4}
\usepackage{amssymb}
\usepackage{amsmath}
\usepackage[dvipsone]{graphics}
\usepackage{epsfig}
\usepackage{bm}
\expandafter\ifx\csname package@font\endcsname\relax\else
\expandafter\expandafter
 \expandafter\usepackage
 \expandafter\expandafter
 \expandafter{\csname package@font\endcsname}%
\fi
\newcommand{\be}{\begin{equation}}
\newcommand{\ee}{\end{equation}}
\newcommand{\ba}{\begin{eqnarray}}
\newcommand{\ea}{\end{eqnarray}}

\renewcommand{\l}{\label}
\newcommand{\f}{\frac}

\renewcommand{\a}{\alpha}
\renewcommand{\b}{\beta}
\newcommand{\p}{\partial}
\renewcommand{\le}{\left}
\renewcommand{\r}{\right}
\renewcommand{\prd}{{\it Phys. Rev. D}}
\renewcommand{\apj}{{\it Astrophys. J.}}
\newcommand{\aj}{{\it Astron. J. (USA)}}
\newcommand{\cqg}{{\it Class. Quant. Grav.}}

\newcommand{\cmda}{{\it Cel. Mech. Dyn. Astron.}}
\newcommand{\mnras}{{\it Mon. Not. Roy. Astron. Soc.}}
\newcommand{\ncb}{{\it Nuovo Cim. B}}
\newcommand{\pla}{{\it Phys. Lett. A}}
\newcommand{\physrep}{{\it Phys. Rep.}}
\newcommand{\rmph}{{\it Rev. Mod. Phys.}}
\newcommand{\lrr}{{\it Liv. Rev. Rel.}}
\newcommand{\jmp}{{\it J. Math. Phys.}}
\newcommand{\sa}{{\it Sov. Astron.}}
\newcommand{\aap}{{\it Astron. Astrophys.}}
\newcommand{\ijmp}{{\it Int. J. Mod. Phys. D}}
\newcommand{\apjl}{{\it Astrophys. J. Lett.}}
\newcommand{\pasp}{{\it Publ. Astron. Soc. Pacific}}
\begin{document}
\title{Gravitational bending of light by planetary multipoles and its measurement with microarcsecond astronomical interferometers}
\author{Sergei M. Kopeikin}
\email{kopeikins@missouri.edu}
\affiliation{Department of Physics \& Astronomy, University of
Missouri-Columbia, Columbia, MO 65211, USA}
\author{Valeri V. Makarov}
\email{vvm@caltech.edu}
\affiliation{Michelson Science Center, California Technology Institute, Pasadena, CA 91125 }
\date{\today}

\begin{abstract}\noindent
General relativistic deflection of light by mass, dipole, and quadrupole moments of gravitational field of a moving massive planet in the Solar system is derived in the approximation of the linearized Einstein equations. All terms of order 1 $\mu$as are taken into account, parametrized, and classified in accordance with their physical origin. The monopolar light-ray deflection, modulated by the radial Doppler effect, is associated with the total mass and radial velocity of the gravitating body. It displaces the apparent positions of stars in the sky plane radially away from the origin of the celestail coordinates associated with the planet. The dipolar deflection of light is due to a translational mismatch of the center of mass of the planet and the origin of the planetary coordinates caused by the inaccuracy of planetary ephemeris. It can also originate from the difference between the null cone for light and that for gravity that is not allowed in general relativity but can exist in some of the alternative theories of gravity. The dipolar gravity field pulls the apparent position of a star in the plane of the sky in both radial and orthoradial directions with respect to the origin of the coordinates. 
The quadrupolar deflection of light is caused by the physical oblateness, $J_2$, of the planet, but in any practical experiment it will have an admixture of the translation-dependent quadrupole due to inaccuracy of planetary ephemeris. This leads to a bias in the estimated value of $J_2$ that should be minimized by applying an iterative data reduction method designed to disentangle the different multipole moments and 
to fit out the translation-dependent dipolar and quadrupolar components of light deflection. The method of microarcsecond interferometric astrometry has the potential of greatly improving the planetary ephemerides, getting unbiased measurements of planetary quadrupoles, and of thoroughly testing the null-cone structure of the gravitational field and the speed of its propagation in the near-zone of the Solar system. We calculate the instantaneous patterns of the light-ray deflections caused by the monopole, the dipole and the quadrupole moments, and derive equations describing apparent motion of the deflected position of the star in the sky plane as the impact parameter of the light ray with respect to the planet changes due to its orbital motion. We discuss the observational
capabilities of the near-future optical (SIM) and radio (SKA) interferometers for detecting the Doppler modulation of the radial deflection, and the dipolar and quadrupolar light-ray bendings by the Jupiter and the Saturn.    
\end{abstract}
\maketitle
\newpage
%\listoffigures
\tableofcontents
\newpage
\section{Introduction}\l{i}
Attaining the level of a microarcsecond ($\mu$as) positional accuracy and better will completely revolutionize fundamental astrometry by merging it with relativistic gravitational physics. Beyond the microarcsecond threshold, one will be able to observe a new range of celestial physical phenomena caused by gravitational waves from the early universe and various localized astronomical sources, space-time topological defects, moving gravitational lenses, time variability of gravitational fields of super-massive binary black holes located in quasars, and many others \cite{kgwinn,ksge,kopmak}. Furthermore, this will allow us to test general theory of relativity in the Solar system in a dynamic regime, that is when the velocity- and acceleration-dependent components of gravitational field (the metric tensor) of the Sun and planets bring about observable relativistic effects in the light deflection, time delay and frequency \cite{kspv}, to an unparalleled degree of precision. 

Preliminary calculations \cite{bkk} reveal that the major planets of the Solar system are sufficiently massive to pull photons by their gravitational fields, which have significant multipolar structures \cite{jpsat}, in contrast with the Sun whose quadrupole moment is only $J_{2\odot}\leq 2.3\times 10^{-7}$ \cite{pj,pit}. Moreover, in the case of a photon propagating near the planet the interaction between the gravitational field and the photon can no longer be considered static, because
the planet moves around the Sun as the photon traverses
through the Solar system
\cite{ks,km}. The optical interferometer designed for the space astrometric mission SIM \cite{sim} is capable of observing optical sources fairly close in the sky projection to planetary limbs with a microarcsecond accuracy. Similar resolution can be achieved for radio sources (quasars) with the Square Kilometer Array (SKA) \cite{ska} if it is included to the inter-continental baseline network of VLBI stations \cite{freid}. The Gaia \cite{gaia} and OBSS \cite{obss} astrometric projects represent another alternative path to microarcsecond astrometry \cite{crosta}.
It is a challenge for the SIM and SKA interferometers as well as for Gaia and OBSS to measure the gravitational bending of light caused by various planetary multipoles and the orbital motion of the planets. This measurement, if sussessful, will be a cornerstone step in further deployment of theoretical principles of general relativity to fundamental astrometry and navigation at a new, exciting technological level.  

The first detection of gravitational bending of light by Jupiter was conducted in 1988 March 21 \cite{tl} (see also \cite{germanteam}), and the deflection term associated with the monopole field of Jupiter was determined to an accuracy of $\simeq 15\%$ to be in agreement with Einstein's general relativity theory. Later on, the Hubble Space Telescope was used to measure the gravitational deflection of light of the bright star HD 148898 as it passed within a few seconds of arc near Jupiter's limb on 24 September 1995 \cite{hst}. Kopeikin \cite{kapjl} proposed to use Jupiter's orbital motion in order to measure the retardation effect in the time of transmission of the gravitational perturbation from the Jupiter to the photon, that appears as a small excess to the Shapiro time delay caused by the change in the gravitational force due to the motion of the planet. This proposal was executed experimentally in 2002 September 8, and the retardation of gravity (as compared with the time of propagation of light from the Jupiter to the observer) was measured to $\simeq 20\%$  accuracy \cite{fk}. The retardation of gravity is determined by the ultimate value of the speed of gravity that was found to be equal to the speed of light as implied by general relativity. Physical interpretation of the gravity retardation effect crucially depends on the way of understanding of the theoretical concepts of general relativity and on how these concepts and principles are applied in experimental gravitational physics. Different authors elaborate on the multi-faceted realizations of these principles and depending on the approach tend to emphasize various physical aspects of the jovian experiment \cite{kf,kf-pla,ni,will-livr,k-ijmp,arak,scb,ser}. 

The jovian experiment of 2002 stimulated researchers from the Gaia consortsium \cite{gaia} to take a next step in exploring the gravitational bending of light by major planets. Most notably, Crosta \& Mignard \cite{cm} proposed to measure the deflection-of-light term associated with the axisymmetric (quadrupolar) part of Jupiter's gravitational field \cite{ander,titan}. Their work is  
aimed at converting the earlier theoretical calculations \cite{kl-azh,k-jmp,tey} of light bending by gravitational multipoles into a practical algorithm for Gaia, thus, extending the relativistic techniques of astrometric data reduction worked out in a number of previous papers \cite{ks,km,kk,kl,cdf1,cdf2}. Detection and precise measurement of the quadrupolar deflection of light in the Solar system is important for providing an independent experimental support for the theory of gravitational lensing by clusters of galaxies. The cluster's gravitational potential (including invisible dark matter) is reconstructed from the observed distortion of images of background quasars under the assumption that the multipolar field of the gravitational lens deflects light exactly as predcited by general relativity \cite{css,fa}. We emphasize that this assumption is highly plausible but still pending experimental confirmation that might be crucial for getting an unbiased estimate of the amount of dark matter in the universe. 

The work \cite{cm} is stimulating but it is incomplete and should be extended in several respects. First, it assumes that light propagates in the field of a static planet while the Jupiter moves on its orbit as light traverses the Solar system toward the observer. The authors of \cite{cm} do not specify at what particular instant of time the planet is fixed on its orbit and whether the orbital velocity of the Jupiter is essential for proper interpretation of the microarcsecond observation of light deflection. Second, Crosta \& Mignard \cite{cm} implicitly assume that the center of mass of the planet deflecting light rays coincides precisely with the origin of the inertial coordinate system in the sky used for interpretation of the apparent displacements from the gravity-unperturbed (catalogue) positions of stars. This makes the dipole moment, $I^i$, of the gravitational field of Jupiter vanish, which significantly simplifies the theoretical calculation of light bending. However, the assumption of $I^i=0$ is not realistic because the instantaneous position of the planet's center of mass on its orbit is known with some error due to the finite precision of the jovian ephemeris limited to a few hundred kilometers \cite{pit}. The ephemeris error will unavoidably bring about a non-zero dipole moment that must be included in the multipolar expansion of the gravitational field of the planet along with its mass and the quadrupole moment. In other words, any realistic set of measurements can only be adequately interpreted within a certain model of the relativistic deflection of light, which includes parameters accounting for a possible shift of the true position of the planetary center of mass from the assumed origin of the deflection pattern, and their relative motion.
It is important to realize that introduction of a static local coordinate system linked to the global star reference frame at few microarcsecond level, will
give rise to non-physical deflection patterns caused by the planetary motion, which must be sorted out in any realistic experimental setup.
 
The dipolar anisotropy in the light-ray deflection pattern is a spurious, coordinate-dependent effect and, hence, should be properly evaluated and supressed as much as possible by fitting the origin of the coordinate system used for data analysis to the center of mass of the planet. Until the effect of the  gravitational dipole is properly removed from observations it will forge a model-dependent quadrupolar deflection of light because of the translational change in the planetary moments of inertia -- the effect known as the parallel-axis or Steiner's theorem \cite{arn} (see Eq. (\ref{9}) below). This translationally-induced quadrupolar distortion of the light-ray deflection pattern should be clearly discerned from that caused by the physical quadrupole moment of the planet $J_2$. This paper discusses the theoretical and observational aspects of monopolar, dipolar, and quadrupolar light-ray deflections. We investigate how the spurious deflections can be separated from the physical ones caused by the {\it intrinsic} quadrupole moment of the planet $J_2$, and how to use the measurement of the dipolar anisotropy to test the relativistic effects caused by the time-dependent component of the gravitational field deflecting light rays. 

We describe our mathematical notations in section \ref{not}. Propagation of light ray in flat space-time and parametrization of the gravity-unperturbed light-ray trajectory is given in section \ref{ulrt}. Analytic description of the retarded gravitational field of a moving planet and its multipolar decomposition is introduced in section \ref{grf}. Null geodesic equations describing the gravitationally-perturbed light-ray propagation, mathematical technique of their integration, and definition of the observed angle of the gravitational light-ray deflection are included to section \ref{dan}. Specific light-ray deflection patterns caused by mass, dipole, and quadrupole moments of gravitational field are outlined in section \ref{lrdp}. We discuss observable relativistic effects in section \ref{ore}, and technological capabilities of SIM and SKA interferometers to pursue astrometric measurements of these effects with microarcsecond accuracy, in section \ref{ocap}. 

\section{Notations}\l{not}

To simplify our theoretical equations we use the system of units in which the fundamental speed $c$ and the universal gravitational constant $G$ are both equal to unity: $G=c=1$. If necessary, these fundamental constants can be restored in equations by making use of dimensional arguments \cite{mtw,LL}.

We further assume that general relativity is valid, which implies that gravity operates on the null cone and the force of gravitational interaction propagates with the fundamental speed $c$ \cite{low}. We suggest that each photon incoming to the Solar system propagates in vacuum, and its physical speed is equal to $c$. It simplifies our consideration as we avoid discussing the perturbations in propagation of light due to dispersive effects in plasma.

Latin indices take values 1,2,3, and the Greek ones run from 0 to 3. The Kroneker symbol (a diagonal unit matrix) is denoted $\delta_{ij}={\rm diag}(1,1,1)$, and the fully anti-symmetric symbol of Levi-Civita $\epsilon_{ijk}$ is defined in such a way that $\epsilon_{123}=+1$. The Minkowski metric is $\eta_{\a\b}={\rm diag}(-1,1,1,1)$. Greek indices are raised and lowered with the Minkowski metric $\eta_{\a\b}$. By convention, Latin indices are raised and lowered with the Kroneker symbol $\delta_{ij}$ which makes no difference between super- and sub-script Latin indices. Repeated indices indicate the Einstein summation rule, for example, $\eta_{\a\b}A^\a B^\b=A_\b B^\b=A_0B^0+A_1B^1+A_2B^2+A_3B^3$.

An arrow above a letter denotes a spatial vector, for instance ${\vec x}=x^i=(x^1,x^2,x^3)$. A dot or a cross between two spatial vectors denote the Euclidean scalar or vector products respectively: ${\vec A}\cdot{\vec B}=A_iB^i=A_iB_i=A^iB^i$, and ${\vec A}\times{\vec B}=({\vec A}\times{\vec B})^i=\epsilon_{ijk}A^jB^k$. Angular brackets around a pair of the Latin indices of a spatial tensor of rank two denote its symmetric and trace-free (STF) part, for example,  \cite{th}
\be\l{stf} 
I^{<ij>}\equiv\f12\le(I^{ij}+I^{ji}\r)-\f13\delta^{ij}I^{kk}\;,\ee 
where $I^{kk}=\delta^{kp}I^{kp}$.

Partial derivatives with respect to four-dimentional coordinates $x^\a$ are denoted with comma so that for any differentiable function $F(t,{\vec x})$ one has $F_{,\a}\equiv \p F/\p x^\a$. Partial derivatives of $F(t,{\vec x})$ with respect to spatial coordinates $x^i$ are denoted as $F_{,i}\equiv\p F/\p x^i\equiv{\vec\nabla}F$. Partial derivative of function $F(t,{\vec x})$ with respect to time is denoted $F_{,0}\equiv\p F/\p t$.
Total derivative of $F(t,{\vec x})$ with respect to time is denoted with overdot, that is \be\dot F\equiv\f{dF}{dt}=\f{\p F}{\p t}+\f{d\vec{x}}{dt}\cdot{\vec\nabla}F\;.\ee Notice that, in general, $\dot F\not=\p F/\p t$ except when $F$ is a function of time only.

\section{The Unperturbed Light-ray Trajectory}\l{ulrt}

Let us introduce a global coordinate system $x^\a=(x^0,x^i)=(t,x^i)$ which coincides with an inertial frame at infinity where the gravitational field is absent. We shall assume that the coordinate system $x^\a$ is static with respect to the observer but we shall prove in section \ref{dan} that our results are gauge-independent and Lorentz-invariant and, in fact, are valid in any other frame moving with respect to the observer with constant velocity.

\begin{figure*}
\includegraphics{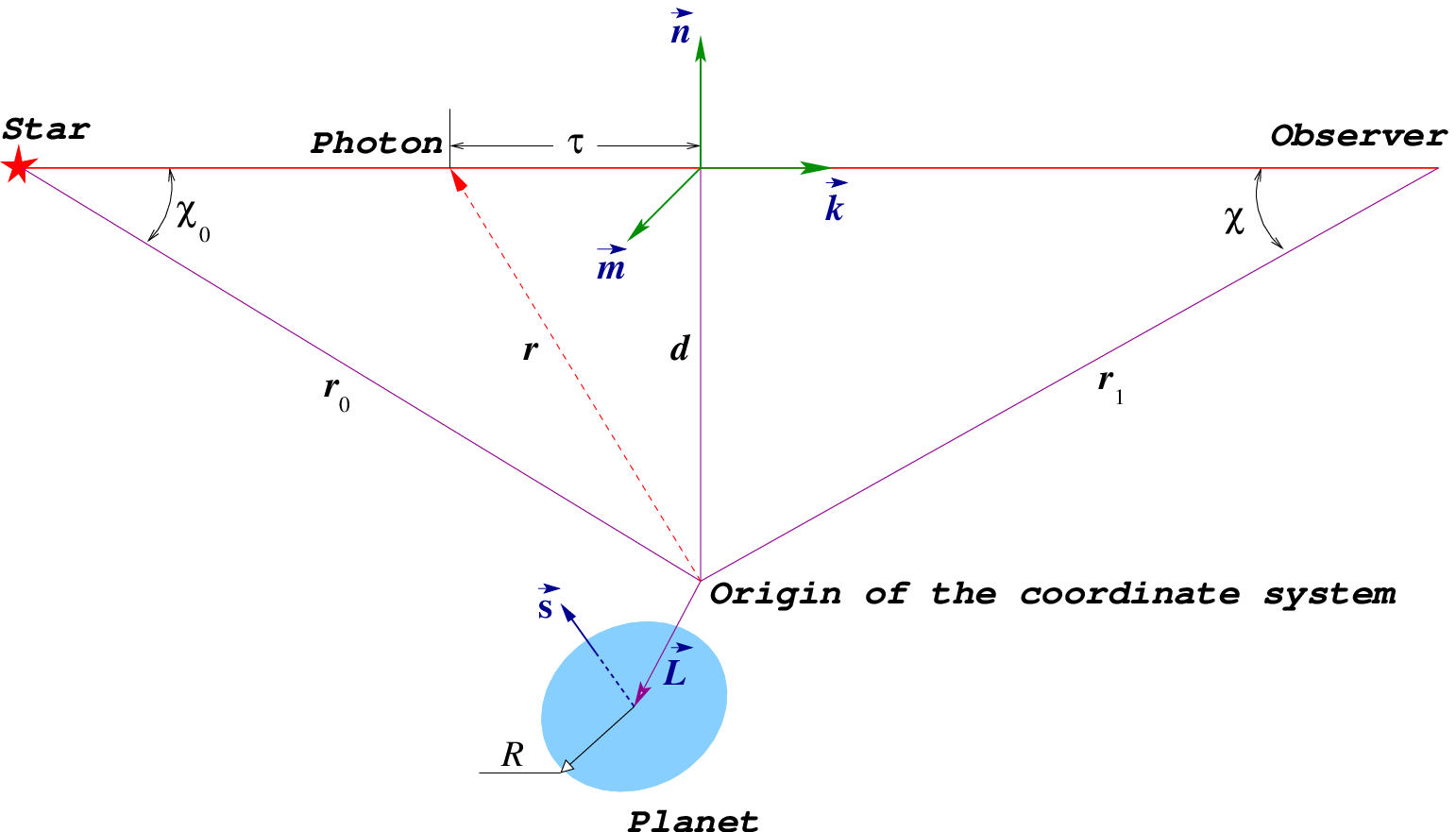}
\caption[Propagation of light]{\label{fig_1}
Light propagates from star to observer in the direction shown by the unit vector ${\vec k}$. The planet is displaced from the origin of the coordinate system by vector $\vec{L}$ that is a parameter of the data analysis algorithm. Its rotation axis is specified by the unit vector ${\vec s}$. The impact parameter of the light ray is $d$. Two unit vectors, ${\vec n}$ and ${\vec m}$, are orthogonal to the vector ${\vec k}$ and form a "plane of the sky" that is perpendicular to the line of sight. $\chi$ denotes the angle between the star and the assumed position of the planet in the coordinate origin.}
\end{figure*}

Let us consider a bundle of light-rays emitted by a source of light (star, quasar) simultaneously and moving as a narrow beam along parallel lines toward the Solar system. In the absence of gravitational field each light particle (photon) from the bundle propagates in the coordinate system $x^\a$ along a straight line
\be\l{1}
x^i=x^i_0+k^i(t-t_0)\;,
\ee
where $t_0$ and $x^i_0=x^i(t_0)$ are the time and space coordinates of the photon at the time of emission, and $k^i$ is the unit vector along the unperturbed photon's trajectory (see Fig. \ref{fig_1}). We assume that the photon hits the detector (is observed) at time $t_1$ when its coordinate $x^i_1=x^i(t_1)$. Let us denote the time of the closest approach of the photon to the origin of the coordinate system as \cite{not}
\be\l{oku}t_*=t_0-{\vec k}\cdot{\vec x}_0\;.\ee  
It is mathematically convenient \cite{k-jmp} to introduce a parameter $\tau$ along the unperturbed light ray \be\l{2}
\tau=t-t_*\;,
\ee
and a constant (light-ray impact-parameter) vector 
\be\l{3}
\xi^i=P^{ij}x^j=P^{ij}x^j_0\;,
\ee
that points out from the origin of the coordinate system to the point of the closest approach of the unperturbed light ray (see Fig. \ref{fig_1}). Here $P^{ij}=\delta^{ij}-k^ik^j$ is the operator of projection onto the plane of the sky, that is orthogonal to the vector $k^i$. By definition, $P^{ij}P^{jk}=P^{ik}$.
Parametrization of the unperturbed light-ray trajectory given by Eqs. (\ref{oku})--(\ref{3}) converts Eq. (\ref{1}) to 
\be\l{pj}
x^i=k^i\tau+\xi^i\;,
\ee
where $\tau$ and $\xi^i$ are independent of each other, and can be considered as coordinates in the 2+1 manifold of the light-ray bundle because the projection (\ref{3}) makes $\xi^i$ have only two independent components. 

At each instant of time $t$, the distance $r$ of the photon from the origin of the coordinate system is
\be\l{7}
r=\sqrt{\tau^2+d^2}\;,
\ee
where $d=|{\vec\xi}|$ is the absolute value of the impact parameter of the photon, that is constant for each light ray from the parallel beam of the light-ray bundle moving in the direction $\vec k$. Notice that definition (\ref{2}) implies that  $\tau=0$ when $t=t_*$, $\tau <0$ when $t<t_*$, and positive otherwise (see Fig. \ref{fig_1}).

\section{The Gravitational Field}\l{grf}
\subsection{The field equations}\l{feq}
We shall consider the Solar system to be isolated and space-time to be  asymptotically flat which means there are no other masses outside of the Solar system. The gravitational field of the Solar system is produced by the Sun and the planets which curve space-time and deflect light. In what follows, for the sake of simplicity we are taking into account the gravitational field of two bodies only - the Sun and a planet moving around the barycenter of the Solar system. Futhermore, we consider the solar gravitational field as spherically-symmetric in its own, proper reference frame because in this frame the dipole and quadrupole moments of the Sun are negligible. Gravitational deflection of light by the Sun is well-known \cite{br} and we will basically focus on the discussion of the light bending by a moving axisymmetric planet.

The planet moves around the barycenter of the system  as a light ray propagates from a star toward the observer. Position of the planet with respect to the origin of the coordinates $x^\a$ at time $t$ is defined by vector $\vec x_P=\vec x_P(t)$, its velocity is denoted $\vec v_P=d\vec{x}_P/dt$, and acceleration $\vec a_P=d\vec{v}_P/dt$. We assume that the planet is axisymmetric around the unit vector $\vec s=\vec s(t)$ that defines the rotational axis of the planet at time $t$. This vector can change its orientation in space due to precession. Our calculation method is general enough, and we do not need to assume that the parameters characterizing translational and rotational motion of the light-ray deflecting bodies are either constant or equal to zero, as it was postulated in \cite{cm} where the authors assumed that $\vec x_P=0$, along with $\vec v_P=\vec a_P=0$ and $\vec s$ is a constant vector. 
We shall calculate relativistic deflection of light by a planetary axisymmetric gravitational field but ignore the relativistic effects  caused by the gravimagnetic field due to the {\it intrinsic} rotation of planet, since it is negligibly small and can not be detected at the microarcsecond resolution \cite{km}. We refer the reader to paper \cite{bkk} where detailed estimates of the magnitude of the gravitational light deflection caused by various parameters of the Solar system bodies are given in Table 1.

The gravitational field of the Solar system is described by the metric tensor 
\be\l{mt} 
g_{\a\b}=\eta_{\a\b}+h_{\a\b}\;,
\ee 
where $\eta_{\a\b}$ is the constant Minkowski metric, and $h_{\a\b}=h_{\a\b}(t,{\vec x})$ is its perturbation which is associated in general relativity with gravitational potentials. Let us impose the harmonic gauge condition on the potentials 
\be\l{gc} 
h^{\a\b}{}_{,\b}-\f12\eta^{\a\b}h_{,\b}=0\;,\ee 
which is convenient and mathematically powerful choice for solving Einstein's gravity field equations \cite{fock,fock1}. Outside of the planet, in vacuum, and in the harmonic gauge (\ref{gc}), the linearized Einstein equations for the field $h_{\a\b}$ are homogeneous wave equation \cite{th,bd}
\be\l{he}
\le(-\f{\p^2}{\p t^2}+{\vec\nabla}^2\r)h_{\a\b}=0\;.
\ee
A question can arise about universality of the harmonic gauge and physical interpretation of the wave solution of the gravity-field equation (\ref{he}). Indeed, Einstein's theory of general relativity is formulated in a covariant tensor form while the wave equation (\ref{he}) is valid in a particular harmonic gauge only. First of all, we notice that the harmonic gauge is not reduced to a single coordinate system but admits a whole class of both global and local harmonic coordinates related to each other by coordinate transformations which do not violate the harmonic gauge condition (\ref{gc}). The class of the global coordinates consists of the reference frames which are moving with respect to each other with constant velocities in the asymptotically flat space-time. These asymptotically-inertial reference frames are connected thorugh the Lorentz transformation as was shown by Fock \cite{fock1} for a generic case of gravitational field including all non-linearities of Einstein's equations. Local harmonic coordinates were introduced to gravitational physics by Thorne and Hartle \cite{thar}. The law of transformation between the local harmonic coordinates extends the linear Lorentz transformation to the class of harmonic polynomials that are solutions of the homogeneous wave equation \cite{kcelm,bkop,bkop1,dsx1,klvo,kv}. They have a great practical value for modern fundamental astronomy \cite{sof,k06}. Second important observation is that the class of the harmonic coordinates is actually much more generic than physicists used to think. In particular, we have shown \cite{ksge,kkp,kkor} that the standard Arnowitt-Deser-Misner gauge \cite{adm} widely used for dynamic formulation of general relativity, may be viewed as a sub-class of the harmonic gauge suitable for description of gravitational field of an isolated astronomical system in vacuum. Final remark is that we always work in this paper with solutions of the gravity-field wave equation (\ref{he}) in terms of an observable quantity that is the deflection angle of a light ray referred to asymptotically-flat space-time. This quantity is invariant with respect to the gauge transformations and, hence, all our results obtained in this paper are valid in arbitrary gauge. The harmonic gauge is simply a convenient mathematical tool facilitating calculations and physical interpretation of the observed relativistic effects.  

\subsection{The gravitational multipoles}\l{grmul}
A general solution of Eq. (\ref{he}) is given in the form of a multipolar expansion \cite{th,bd} depending on time-dependent parameters characterizing the {\it intrinsic} multipole moments of the planet as well as on the displacement $\vec x_P$ of its center of mass from the origin of the coordinate system. We shall omit in this paper the relativistic effects of the planet's spin, and consider only the mass monopole, dipole, and quadrupole terms in the multipolar expansion of the gravitational field. Relativistic effects caused by the spin and higher-order mass multipoles are discussed in \cite{k-jmp,km,kkp,kkor}.

Because Eq. (\ref{he}) is linear, we can consider the gravitational field of the Solar system to be a linear superposition of the individual fields. For each massive body the solution of Eq. (\ref{he}) is \cite{th,bd}
\ba\l{4}
h_{00}&=&\f{2M}{r}-\f{\p}{\p x^i}\le[\f{2I^i(s)}{r}\r]+\f{\p^2}{\p x^i\p x^j}\le[\f{I^{<ij>}(s)}{r}\r]\;,\\\l{5}
h_{0i}&=&-\f{4\dot I^i(s)}{r}+\f{\p}{\p x^j}\le[\f{2\dot I^{<ij>}(s)}{r}\r]\;,\\\l{6}
h_{ij}&=&\delta_{ij}h_{00}+\f{2\ddot I^{<ij>}(s)}{r}\;,
\ea
where $r$ is the distance from the origin of the coordinate system to the field point $x^\a=(t,x^i)$, and the angular brackets around two indices denote the STF tensor as explained in Eq. (\ref{stf}). Notice that the second time derivative of the dipole moment $I^i$ is absent due to the law of conservation of the linear momentum of the body \cite{th,bd}.

\begin{figure*}
\includegraphics{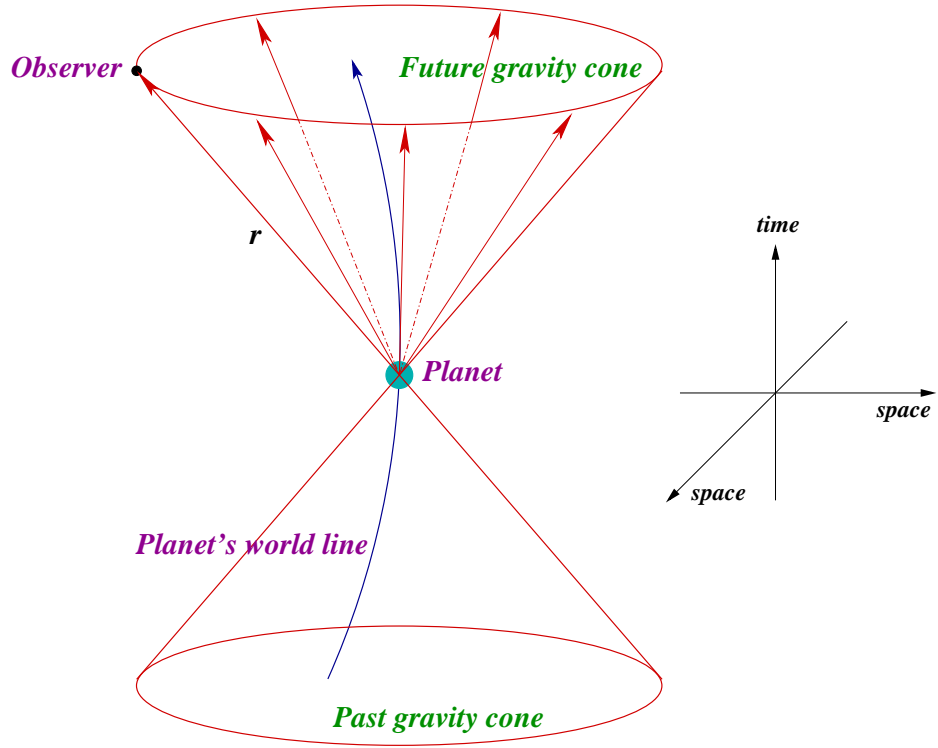}
\caption[Gravity null cone]{\label{fig_gwnc}
Gravitational field of a planet is a retarded solution of the gravity-field wave equation (\ref{he}). In general relativity, gravitational field progresses on the hypersurface of a null cone from the past to the future. Directions of the propagation of the gravitational field of the planet are shown by arrows. The picture assumes that the coordinate-dependent effects associated with the dipole moment $I^i$ are excluded ($I^i=0$). Observer measures the gravitational field at time $t$ when the planet is located at the retarded position on its orbit at the retarded time $s=t-r$, which is a null characteristic of the gravitational field.}
\end{figure*}

The gravitational field described by Eqs. (\ref{4})--(\ref{6}) should be interpreted as that created by a massive planet placed at the origin of the coordinate system and characterized by several parameters, of which $M$ is a constant mass of the planet, $I^i(s)$ is the dipole, and $I^{<ij>}(s)$ is the quadrupole moment -- both taken at the retarded instant of time 
\be\l{rt}
s=t-r\;. 
\ee
The retardation is a direct consequence of the retarded (causal) solution of the gravitational wave equation (\ref{he}). The null cone corresponding to the causal region of influence of the gravitational field of a planet is shown in Fig. \ref{fig_gwnc}. The planet moves along its own world line so that observer is expected to see any effect of the planet's gravitational field with the retardation that is due to the finite speed of the propagation of gravity on the future part of the null cone. Figure  \ref{fig_gwnc} gives a general-relativistic picture of the process of propagation of gravity from the moving planet to the observer but it is also valid in a number of alternative theories of gravity \cite{kf,koni}.   

The dipole and quadrupole moments are fully taken into account because they produce a relativistic deflection of light that can be of significant  value for several planets \cite{bkk,cm}. In general relativity, all relativistic effects due to the dipole moment of the gravitational field are coordinate-dependent and, hence, can be eliminated if the origin of the coordinate system is placed exactly at the center of mass of the real planet. However, we retain the dipole moment $I^i$ in the multipolar decomposition of the planetary gravitational field (\ref{4})--(\ref{6}) because in practice we do not know accurately enough where the center of mass or the planet is and, hence, one needs to parametrize its position. Furthermore, we should consider that even if we were able to to render $I^i=0$ at a particular instant of time, the planet has a residual velocity with respect to the coordinate frame $x^\a$, and its center of mass, $x_P^i$, is not fixed at the origin of this frame for any observation of finite duration. It is therefore, important to  properly parametrize and evaluate the influence of the dipole moment on the light deflection.

If the center of mass of the planet is shifted from the origin of the coordinate system by a spatial distance $L^i=x_P^i$, the dipole and quadrupole moments of the gravitational field entering the metric tensor perturbations (\ref{4})--(\ref{6}) are defined in the linearized approximation of general relativity by the following equations 
\ba\l{8}
I^i&=&Mx_P^i\;,
\\\l{9}
I^{<ij>}&=&J^{<ij>}+Mx_P^{<i}x_P^{j>}\;,
\ea
where $J^{<ij>}$ is the {\it intrinsic} quadrupole moment of the planet in its own proper reference frame whose origin coincides with the planetary center of mass, and we made use of the parallel-axis theorem \cite{arn} to split $I^{ij}$ to $J^{ij}$ and the translational term $Mx_P^{<i}x_P^{j>}$. Equations (\ref{8}), (\ref{9}) approximate more precise, post-Newtonian definitions of the multipole moments used in the relativistic celestial mechanics of N-body system \cite{kv}. However, the post-Newtonian corrections to equations (\ref{8}), (\ref{9}) yield terms in the light deflection which are much less than 1 $\mu$as, and hence, can be ignored in the present paper.  

Both the dipole, $Mx_P^{i}$, and the translational quadrupole, $Mx_P^{<i}x_P^{j>}$, will affect the post-fit results of our light-ray deflection measurement, and thus, are directly observable in the light-ray deflection pattern. But these deflections are spurious, coordinate-dependent effects and the goal of the data analysis system is to reduce their impact on the values of the fitting parameters within the accuracy of astrometric observations. To this end the dipolar and quadrupolar deflections associated with the translation $L^i=x^i_P$ must be incorporated to the data analysis in order to assume full control on their influence on the values of the fitting physical parameters like $J_2$, etc. If this is not done, the coordinate-dependent effects can not be eliminated and the light deflection due to the dipole, $Mx_P^{i}$, and the translational quadrupole, $Mx_P^{<i}x_P^{j>}$, may exceed the physical deflection of light caused by the {\it intrinsic} quadrupole moment, $J_2$, of the planet, thus, making its measurement in the gravitational light-ray deflection experiments impossible.  

In what follows, we assume that the planet is axisymmetric around its rotational axis defined at each instant of time by a unit vector $s^i$. The planet has equal equatorial moments of inertia $A=B$, and the axial moment of inertia $C\not=A$. The dynamic oblateness of the planet is denoted as $J_2=(C-A)/A$ \cite{br,mar}. This definition yields the {\it intrinsic} quadrupole moment represented as an STF tensor of the second rank \cite{mar}
\be\l{j2}
J^{<ij>}=MJ_2R^2\le(s^is^j-\f13\delta^{ij}\r)\;,
\ee
where $R$ is the equatorial radius of the planet. It is immediately seen from Eq. (\ref{j2}) that the quadrupole moment is symmetric and trace-free, that is $J^{<ii>}=0$, in accordance with its definition. We also notice that the first time derivative of the dipole and quadrupole moments
\ba\l{kz}
\dot I^i&=&Mv^i_P\;,\\\l{kz1}
\dot I^{<ij>}&=&2Mx_P^{<i}v_P^{j>}\;,
\ea
should be used in calculations of $h_{0i}$ in Eq. (\ref{5}). We shall neglect the first time derivative of $J^{<ij>}$ in $h_{0i}$, and the second time derivative of the overall quadrupole moment $I^{<ij>}$ in $h_{ij}$ as they lead to higher-order relativistic effects in light deflection. For example, as follows from the subsequent calculations, the light-ray deflections caused by the second derivative of the quadrupole moment $I^{<ij>}$ have magnitude of the order of $\alpha_M(v_P/c)^2$, where $\alpha_M$ is the gravitational bending due to the mass monopole of the planet. In case of Jupiter the maximal value of $\alpha_M\sim 16300$ $\mu$as \cite{bkk,cm}, and $v_P/c\sim 4.5\times 10^{-5}$ \cite{kapjl}, hence, $\alpha_M(v_P/c)^2\ll 1$ $\mu$as. 

\section{The Light-ray Gravitational Perturbations}\l{dan} 
\subsection{The light-ray propagation equation}
The equation of motion of light in gravitational field is given by the light-ray geodesic \cite{mtw,LL}
\be\l{au}
\f{dK^\a}{d\lambda}+\Gamma^\a_{\mu\nu}K^\mu K^\nu=0\;,
\ee
where $K^\a=dx^\a/d\lambda$ is the wave vector, $\lambda$ is the affine parameter along the light ray, and 
\be\l{au0}
\Gamma^\a_{\mu\nu}=\f12 g^{\a\b}\le(g_{\b\mu}{}_{,\nu}+g_{\b\nu}{}_{,\mu}-g_{\mu\nu}{}_{,\b}\r)\;,
\ee
is the Christoffel symbol \cite{mtw,LL}.

When one substitutes the expansion of the metric tensor (\ref{mt}) to Eq. (\ref{au}), and transforms the affine parameter $\lambda$ in this equation to the coordinate time $t$, the light geodesic equation (\ref{au}) is reduced to the following three-dimensional form \cite{br}
\be\l{au1}
\f{d^2x^i}{dt^2}=F^i(t,\vec{x})\;,
\ee
where the gravitational force \cite{br} 
\ba\l{au2}
F^i&=&\f12 h_{00}{}_{,i}-
                          h_{0i}{}_{,0}-
                  \frac{1}2 h_{00}{}_{,0}\dot{x}^i-
                             h_{ik}{}_{,0}\dot{x}^k-
                            ( h_{0i}{}_{,k}-
                               h_{0k}{}_{,i})\dot{x}^k-\\
\nonumber
              &&  h_{00}{}_{,k} \dot{x}^k\dot{x}^i-
            \left( h_{ik}{}_{,j}-
           \frac{1}2 h_{kj}{}_{,i}\right)\dot{x}^k\dot{x}^j+
       \left(\frac{1}2h_{kj}{}_{,0}-
                        h_{0k}{}_{,j}\right)\dot{x}^k\dot{x}^j\dot{x}^i \;,
\ea
depends on the perturbation of the metric tensor $h_{\a\b}$ and the coordinate velocity of photon $\dot x^i$. We notice that making use of time $t$ in place of the parameter $\lambda$ does not change direction of propagation of the light ray, and is merely a technical tool that allows us to express coordinate of the photon as a function of the same time argument which governs evolution of the multipole moments of the gravitational field.    

Equation (\ref{au1}) has an unperturbed ($F^i=0$) solution described in section \ref{ulrt}. In a weak-gravitational field approximation the light-ray geodesic equation (\ref{au1}) has a unique solution given by 
\be\l{elb}
x^i=k^i\tau+\xi^i+\Xi^i(\tau,\vec{\xi}\,)\;,
\ee
where $\Xi^i$ is a small perturbation of the unperturbed light ray trajectory.
In terms of the parameter $\tau$ and the projection coordinates $\xi^i$, equation (\ref{au1}) is significantly simplified, so that the perturbation $\Xi^i$ obeys the following ordinary differential equation \cite{ksge,kkp}
\be\l{10}
\f{d^2\Xi^i}{d\tau^2}=\f12k^\a k^\b\f{\p h_{\a\b}}{\p\xi^i}-\f{d}{d\tau}\le(k^\a h_{i\a}+\f12k^ih_{00}-\f12k^ik^jk^ph_{jp}\r)\;,
\ee
where $k^\a=(1,k^i)$ is a null vector along the unperturbed trajectory of the light ray.
Equation (\ref{10}) describes two relativistic effects -- gravitational bending and time delay of light which are closely related to each other \cite{kkor}. In this paper we discuss only the deflection of light. 

\subsection{The null cone integration technique}
Without any restriction on the ratio of the impact parameter $d$ of the light-ray trajectory to the distance $r_1$ from the origin of the coordinate system to observer, the total angle of the gravitational deflection of light in the plane of the sky is given by a vector \cite{ksge,kk}
\be\l{11}
\alpha^i=-P^{ij}\f{d\Xi^j}{d\tau}+\Delta\alpha^i\;, 
\ee
where $\Delta\alpha^i$ includes relativistic corrections due to both the finite distance $r_0$ of the source of light from the planet, the finite distance $r_1$ between the observer and the planet, and the difference of the local inertial frame of observer \cite{niz,klio2} from the coordinate system $x^\a=(t,x^i)$ introduced for the calculation of light-ray propagation. We further assume that $r_0\rightarrow\infty$, and the observer is at rest at a sufficiently large distance from the light-ray deflecting body which are realistic assumptions \cite{not1,kl-aa}. Under these circumstances $\Delta\alpha^i$ is negligible, and can be omited. Indeed, the largest contribution to $\Delta\alpha^i$ associated with the finite distance $r_0$ of the star from the Solar system (the planet), is smaller than the first term in the left side of equation (\ref{11}) by a factor of $r_1/r_0$ \cite{ksge,ks} that, in case of the Jupiter, is  about $2\times 10^{-5}$ even for the closest star $\alpha$ Centauri, making $\Delta\alpha^i\ll 1$ $\mu$as \cite{nnq}. Contribution to $\Delta\alpha^i$ due to the difference between the local inertial and global coordinate frames is smaller than the first term in the left side of equation (\ref{11}) by a factor of $(d/r_1)(v_P/c)^2$ \cite{ksge,ks} that also amounts to $|\Delta\alpha^i|\ll 1$ $\mu$as, and can be ignored. In cases when $\Delta\alpha^i$ may be important, for example, in studying refraction of light by gravitational waves \cite{ksge,kkor} or in cosmological gravitational lensing where $r_0\not=0$ and $r_1$ is comparable with $r_0$, this term is solved in our papers \cite{ks,km,kkp}.

Integrating Eq. (\ref{10}) along the unperturbed light ray trajectory and substituting the result to Eq. (\ref{11}) yields \cite{ks,km}
\be\l{12}
\alpha^i=-\f12\f{\p }{\p\xi^i}\int^{\tau_1}_{\tau_0}k^\a k^\b h_{\a\b}(\tau,\vec{\xi}\,)d\tau+k^\a P^{ij}\Bigl[ h_{j\a}(s_1,\vec x_1)-h_{j\a}(s_0,\vec x_0)\Bigr]\;, 
\ee
where the integrand is taken on the unpertubed light-ray trajectory, and can be represented as\ba\l{13}
\f12k^\a k^\b h_{\a\b}(\tau,\vec\xi)&=&\f{2M}{r}-\f{\p}{\p\xi^j}\le[\f{2I^j(s)}{r}\r]-\f{d}{d\tau}\le[\f{2k^jI^j(s)}{r}\r]-\f{2k^j\dot I^j(s)}{r}\\\nonumber
&+&\f{\p^2}{\p\xi^j\p\xi^p}\le[\f{I^{<jp>}(s)}{r}\r]+\f{d^2}{d\tau^2}\le[\f{2k^jk^pI^{<jp>}(s)}{r}\r]+\f{\p}{\p\xi^j}\f{d}{d\tau}\le[\f{2k^pI^{<jp>}(s)}{r}\r]\;,
\ea
the limits of integration $\tau_1$ and $\tau_0$ are the values of the parameter $\tau$ taken at the time of observation, $t_1$, and emission, $t_0$, respectively (notice that $\tau_0<0$ and $\tau_1>0$), and $s=t_*+\tau-r$ with $r=r(\tau,\vec\xi\,)$ given by Eq. (\ref{7}). Integration in the right-hand side of Eq. (\ref{12}) can be performed easily if one adopts that in general relativity the speed of propagation of gravity and the speed of light are numerically the same, making the light cone hypersurface be coinciding with one of the gravity null cone, so that in practical experiments gravity interacts with light in a way shown and explained in Fig. \ref{fig_lgwnc}. In some of the alternative theories of gravity the hypersurfaces of the light and gravity null cones do not coincide as the speed of propagation of gravity and light are different \cite{will-livr,kop-cqg,kostel,matt,will-b}. We have discussed the gravitational deflection of light caused by moving bodies, in such theories in a number of other papers \cite{kf,koni,kop-cqg} which can be used as references. We do not discuss the deflection of light in the alternative theories of gravity in the present paper. 

We have found \cite{ksge,kkp,kkor} that for any smooth function of the retarded time, $F(s)/r$, where $r=\sqrt{\tau^2+d^2}$ and $s=t-r=t_*+\tau-\sqrt{\tau^2+d^2}$, the integration can be performed explicitly, and it yields 
\ba\l{14}
\int^{\tau_1}_{\tau_0}\f{\p }{\p\xi^i}\le[\f{F(s)}{r}\r]d\tau&=&\f{\p }{\p\xi^i}\int^{\tau_1}_{\tau_0}\f{F(s)}{r}d\tau=
-\Bigl[(1+\cos\chi)F(s_1)-(1-\cos\chi_0)F(s_0)\Bigr]\f{\xi^i}{d^2}\;,\\\l{14a}
\int^{\tau_1}_{\tau_0}\f{d }{d\tau}\le[\f{F(s)}{r}\r]d\tau&=&\f{F(s_1)}{r_1}-\f{F(s_0)}{r_0}\;,
\ea
where the retarded times 
\ba\l{rta}
s_0&=&t_0-r_0\;,\\\l{rtb} 
s_1&=&t_1-r_1\;,
\ea
$\cos\chi_0=|\tau_0|/r_0$, $\cos\chi=|\tau_1|/r_1$ are referred to distances $r_0=|{\vec x}(t_0)|$, $r_1=|{\vec x}(t_1)|$ of the photon from the origin of the coordinate system taken at the times of emission and observation of light respectively (see Fig. 1). Signs in the right side of Eq. (\ref{14}) are valid in case shown in Fig. \ref{fig_1} when the light-ray deflecting body is located between the source of light and observer. If the observer is located between the source of light and the light-ray deflecting body, Eq. (\ref{14}) is still applicable after replacing $\cos\chi\rightarrow -\cos\chi$ \cite{ksge}. Finally, if $d=r_1$, that is the light ray arrives from the direction perpendicular to that to the light-ray deflecting body, we must take $\cos\chi=0$ in Eq. (\ref{14}). All these situations are carefully discussed in \cite{ksge} to which the reader is referred for a comprehensive theoretical review. In this paper we shall discuss the configuration shown in Fig. \ref{fig_1} because only this case is practically important for the gravitational deflection of starlight by the Solar system planets.

It is noted that the retarded time Eq. (\ref{rtb}) describes the null direction connecting the planet and the observer and lying on the future gravity null cone with the planet at its vertex as shown in Fig. \ref{fig_lgwnc}. Light moves along a different null direction connecting the star and the observer and lying on the past light cone of the observer. Therefore, Eq. (\ref{rtb}) describes the retardation effect in propagation of gravity force from the moving planet to a photon as the photon propagates toward the  observer and is subsequently detected at time $t_1$. This retarded component in the interaction of gravity with light was measured within 20\% in the jovian VLBI experiment on September 8, 2002, and led to the direct observational confirmation of the general relativistic postulate that the speed of gravity and light are the same within the observational error \cite{kapjl,fk,kf}. 

\subsection{The speed of gravity, causality, and the principle of equivalence}\l{sgpe}

\begin{figure*}
\includegraphics{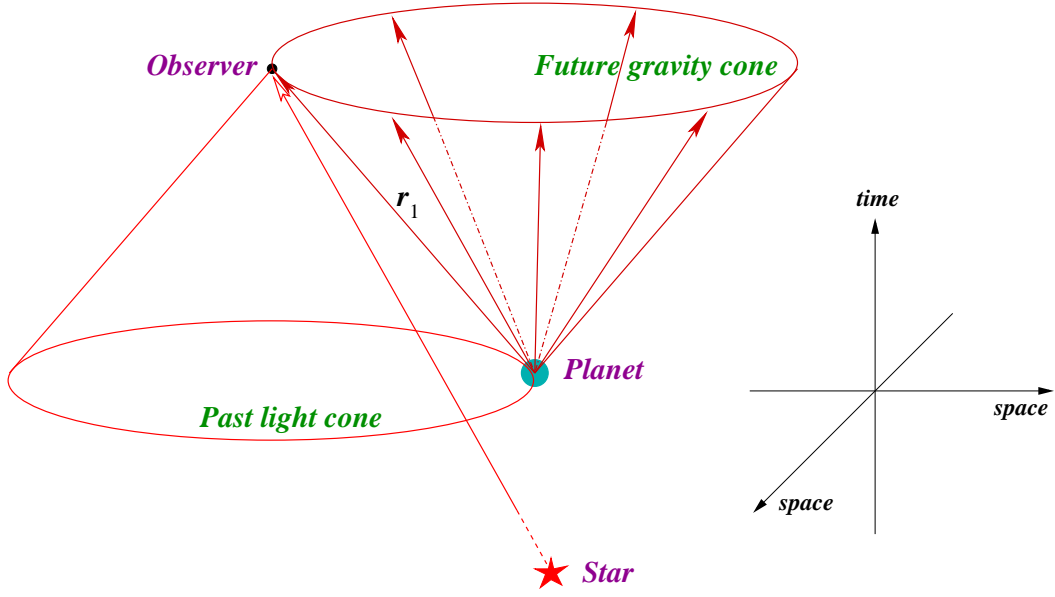}
\caption[Interacting gravity and light null cones]{\label{fig_lgwnc}
Gravitational field of a planet (the case of dipole $I^i=0$ is shown) affects only the particles lying on the hypersurface of the future gravity null cone. A photon emited by a star at time $t_0$ arrives to observer at time $t_1$ along a null direction of the past light cone with a vertex at the observer. Therefore, the future gravity null cone of the planet and the past light cone of the observer must coincide along the null direction that is a null characteristic of the retarded solution of the gravity-field wave equation (\ref{he}). The photon detected at time $t_1$, is deflected by planet's gravity force from the planet's retarded position taken at time $s_1=t_1-r_1$. This effect of the retardation of gravity can be observed by measuring the amount of gravitational deflection of light by a moving planet, and used to determine the speed of gravity field propagating from the planet to the point of observation (see section \ref{dan} and papers \cite{fk,kf}). Note that the retardation of gravity would
not be measurable if the planet were at rest with respect to the observer.}
\end{figure*}

Review article \cite{will-livr} argues that the speed of propagation of the gravity force from the planet to the traversing photon is irrelevant in the light-ray deflection experiments done in the Solar system with a uniformly moving planet, and that the light deflection angle can be calculated correctly even if the speed of gravity is infinite. In other words, the author of \cite{will-livr} mantains the point that the causal structure of the future null cone of the gravity field (see Fig. \ref{fig_lgwnc}) is not essential to the calculation of the light-ray deflection angle so far as acceleration of the planet is ignored. This point of view stems from the belief, explicitly formulated and adopted in the PPN formalism \cite{will-b}, according to which the principle of equivalence demands nothing about the speed of gravity that determines the causal space-time hypersurface on which gravity propagates. We are told that this is because the principle of equivalence operates with derivatives of the metric tensor of the first order (the Christoffel symbols) while the speed of gravity is determined only by the second time derivatives of the metric that appear in the field equations of each metric theory of gravity  (for more detail see section 10.1 in \cite{will-b}). 

This PPN postulate disentangles gravitational field from the geometry of the space-time manifold and ignores existence of other (global or geometric) techniques \cite{mtw} to study the causality of gravitational field which is a direct consequence of its ultimate speed of propagation. The causal structure of gravitational field is derived not so much from the field equations but from exploration of behavior of a set of future directed timelike and/or null geodesics in a given space-time manifold \cite{hawe}. This behavior can be determined, at least in a close neigborhood of any event, by the geodesic equations without imposing Einstein's field equations \cite{wald}. The null geodesics define the causal past of observer, that is the region bounded by the past light cone in Fig. \ref{fig_lgwnc}, and the causal future of the gravitational field of a massive body, that is the region inside the future gravity cone in Fig. \ref{fig_lgwnc}. General-relativistic gravitational interaction of photon with the body implies that the causal past of the observer must touch the causal future of the gravitational field along a null direction as demonstrated in Fig. \ref{fig_lgwnc} \cite{mnot}. 

\begin{figure*}
\includegraphics{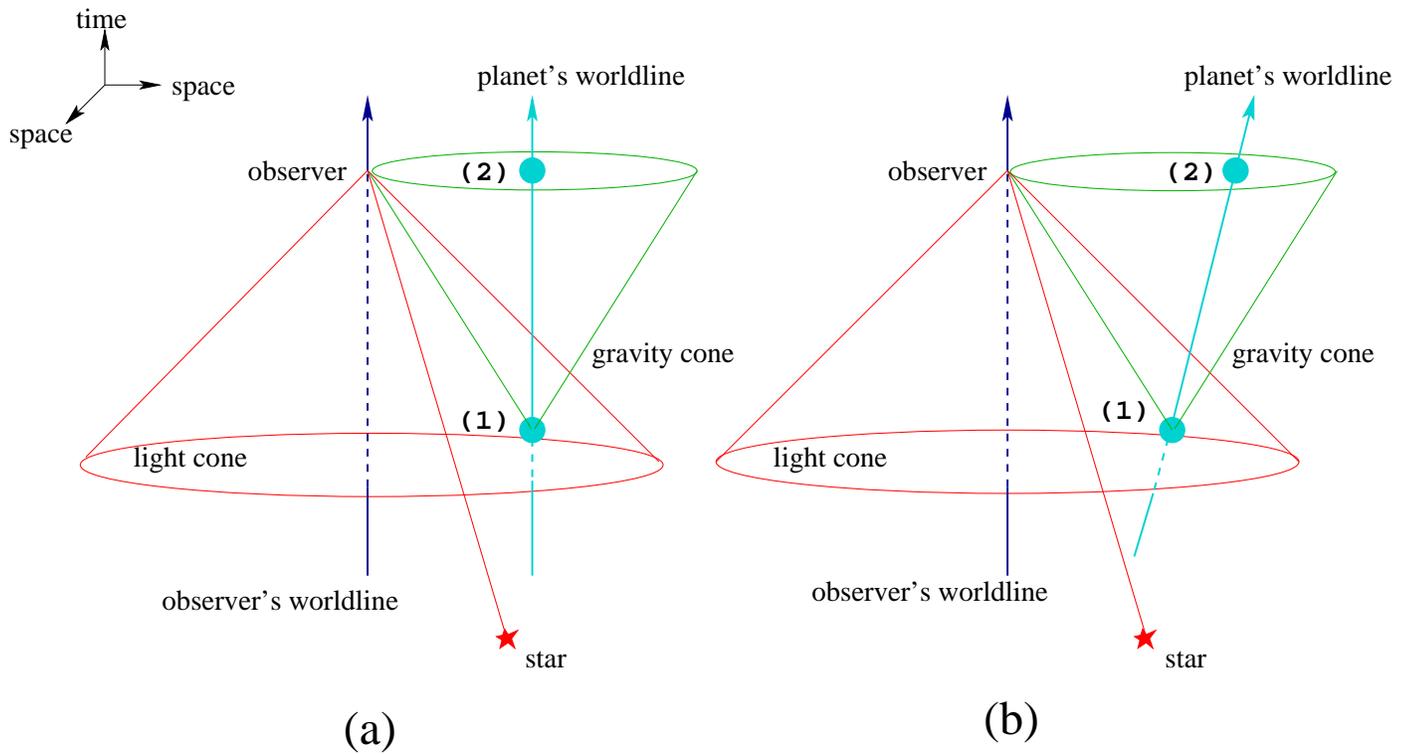}
\caption[Static versus dynamic light-ray deflection]{\label{g_l}
Light-ray deflection by a static (a) and moving (b) planet. In case (a) the distance between the planet and observer does not change as light propagates. Thus, measuring the deflection of light does not allow us to determine experimentally whether the gravity force of the planet acts with retardation from position (1), or instantaneously from position (2). In case (b) the distance between the planet and the observer varies as the photon travels toward observer. The retarded interaction of gravity with light becomes apparent since measuring the angle of the gravitational deflection of light allows us to distinguish between positions (1) and (2) of the planet on its world-line making the causal structure of the gravity cone clearly visible and measurable.}  
\end{figure*}

The equivalence principle tells us that in a local reference frame light moves along a straight line \cite{mtw,LL}. This implies that in the global reference frame the light-ray trajectory is bent because the local reference frame falls in the gravitational field of a global frame with acceleration, and the space is curved \cite{mtw}.  Testing gravitational light-ray deflection is not an experiment in a single local frame of reference as photon propagates through a continuous sequence of such local frames, thus, accumulating the pointwise influence of the gravitational field at different parts of the light-ray trajectory. Photon's wave-vector propagator is an integral of the affine connection that is the gravity force $F^i$ shown in the right-hand side of Eq. (\ref{au1}). This propagator contracted with the projection operator $P^{ij}=\delta^{ij}-k^ik^j$, yields the integrated deflection angle $\a^i$. The magnitude and/or direction of the deflection angle are functions not only of the wave vector of observed starlight, but of the position of the light-ray deflecting body with respect to observer as well. If one is able to derive position of the massive body from the precise measurement of the angle $\a^i$ and to confirm that the body and the observer are connected by a null line which is a characteristic of the future gravity null cone, it gives a direct proof of the causal nature of the gravitational field and allows us to measure its finite speed of propagation \cite{kapjl}. However, this measurement of the causal property of gravity is impossible if the gravitational field is static. Indeed, in case of a static planet the gravitational field does not change as photon moves from star to observer, and the causal character of the gravitational field gets hidden from observer because the planet is always at a fixed distance from the observer (see Fig. \ref{g_l}a). 

The situation changes dramatically if the planet moves with respect to observer because it makes the distance between the planet and the observer depending on time (see Fig. \ref{g_l}b). In this case, photon traverses through the gravitational field that changes on the light-ray trajectory due to the planetary motion, even if this motion is uniform. Had the speed of the propagation of the planet's gravitational field been different from the speed of light it would have unavoidably led to violation of the causal nature of gravity that would be inconsistent with the principle of equivalence \cite{matt}. Indeed, in this case the light null cone would be different from the gravity null cone, and the moving planet would not deflect the observed photon from the retarded position as predicted by general theory of relativity which predicts that the angle of the gravitational deflection of light is a gauge-invariant quantity (compare Fig. \ref{g_l}a with Fig. \ref{g_l}b which are connected via Lorentz transformation). For example, the instantaneous propagation of gravity would imply that one could determine current (as opposed to retarded) position of the planet on its orbit from observation of the gravitational deflection of light, that is the gravitational field would transmit information about the planet's spatial location to observer faster than light. This violates the principle of causality, and we conclude that correct description of the gravitational physics of the light-ray deflection experiment with a moving planet requires taking into account both the light and gravity null cones as demonstrated in Figs. \ref{fig_gwnc}, \ref{fig_lgwnc}, \ref{g_l}, and supported by calculations in this paper as well as the discussion given in \cite{kop-cqg}. Additional arguments backing up the concept of the retardation of gravity in the light-ray deflection experiments by moving planets are discussed in our paper \cite{kf} both in the framework of general relativity and in a bi-metric theory of gravity proposed by Carlip \cite{carla}. The present paper and calculations given in \cite{kf} refute the superficial understanding of gravitational physics of the light-ray deflection experiments advocated in the PPN formalism \cite{will-livr} and in \cite{carla}.
The principle of equivalence does imply the principle of causality for gravitational field and demand the ultimate speed of gravity to be equal to the speed of light which can be (and was) confirmed in the Solar system experiments of starlight deflection by a moving planet \cite{fk}.

\section{Light-ray Deflection Patterns}\l{lrdp}
\subsection{The deflection angle}\l{defa}

In calculation of the light-ray deflection angle all terms related to the time of emission are negligibly small because $1-\cos\chi_0\simeq d^2/(2r_0^2)\ll 1$, and can be omitted since we assume that $r_0\rightarrow\infty$. We also assume that only the stars that are sufficiently close to the planetary limb in the plane of the sky are observed. This makes the angle $\chi\ll 1$, and hence $\cos\chi\simeq 1-d^2/2r_1$ and $\sin\chi=d/r_1\ll 1$. We shall  neglect all terms proportional to $\sin\chi$ and approximate $\cos\chi= 1$ everywhere but in the calculation of the monopolar deflection, where more exact approximation of $\cos\chi$ is required. 

Integrating Eq. (\ref{12}) with the technique shown in Eqs. (\ref{14})--(\ref{14a}), and keeping only the leading terms, we deduce
\be\l{15}
\a^i=\alpha_\odot+\a^i_M+\a^i_D+\a^i_Q\;,
\ee 
where $\a^i_M$, $\a^i_D$, $\a^i_Q$ are the angles of the relativistic deflection of light caused by the planetary mass monopole, dipole, and quadrupole moments respectively, and $\alpha_\odot$ stands for the well-known deflection angle caused by the Sun's gravity \cite{br}. The deflection terms are defined by the following equations
\ba\l{16}
\a^i_M&=&2M(1+\cos\chi)\f{\xi^i}{d^2}\;,\\\l{17}
\a^i_D&=&-2k^j\dot I^j(s_1)(1+\cos\chi)\f{\xi^i}{d^2}-2\f{\p }{\p\xi^i}\le[(1+\cos\chi)\f{\xi^j I^j(s_1)}{d^2}\r]\;,\\\l{18}
\a^i_Q&=&\f{\p^2 }{\p\xi^i\p\xi^j}\le[(1+\cos\chi)\f{\xi^p I^{<jp>}(s_1)}{d^2}\r]\;,
\ea
of which Eq. (\ref{16}) is exact, and in Eqs. (\ref{17}), (\ref{18}) the terms proportional to $\sin\chi$ and higher, have been omitted. These residual terms are of the order of $\a_D d/r_1$ and $\a_Q d/r_1$, that is negligible, since \rm{max}$[\a_D,\a_Q]\simeq 2000$ $\mu$as (see Figs. \ref{fig_3}, \ref{fig_4}), and $d/r_1\simeq 10^{-4}$ for Jupiter and about $10^{-5}$ for Saturn.  

Taking all the partial derivatives with accounting for 
\be\l{19}
\f{\p\cos\chi}{\p\xi^i}=-\f{\xi^i}{d^2}\,\cos\chi\sin^2\chi\simeq\f{d}{r_1^2}\ll \f{1}{d}\;,
\ee
and approximating $\cos\chi=1$, $\sin\chi=0$ in the dipolar and the quadrupolar deflections, we reduce Eqs. (\ref{16})--(\ref{18}) to
\ba\l{20}
\a^i_M&=&\f{4M}{d}(1+\cos\chi)n^i\;,\\\l{dm}
\a^i_D&=&\f{4I^j(s_1)}{d^2}\le(n^in^j-m^im^j\r)-\f{4k^j\dot I^j(s_1)}{d}n^i\;,\\\l{21}
\a^i_Q&=&\f{4I^{<jp>}(s_1)}{d^3}\le(n^in^jn^p-n^im^jm^p-m^im^jn^p-m^im^pn^j\r)\;,
\ea
where $I^i$ and $I^{ij}$ are determined by Eqs. (\ref{8}) and (\ref{9}) respectively, the unit vectors $n^i=\xi^i/d$, $m^i=({\vec k}\times{\vec n})^i$, and again all terms proportional to $\sin\chi$ have been neglected. On the other hand, the term with $\cos\chi$ in Eq. (\ref{20}) must be retained  as it reach magnitude of the order of 2 $\mu$as when $\chi=90^o$ for Jupiter, and is about 1 $\mu$as when $\chi=16^o$ for Saturn, that is the small-angle approximation in Eq. (\ref{20}) is invalid.

It is evident that the deflection angles $\a_D^i$ and $\a^i_Q$ depend on the value of the dipole and quadrupole moments of the planet taken at the retarded time $s_1=t_1-r_1$, where $t_1$ is the time of observation. The deflections caused by the translation $x^i_P$ are unphysical and can be removed, at least in theory, by chosing the origin of the coordinate system used for calculation of the deflection angles at the center of mass of the planet taken at the retarded instant of time with respect to the observer. This situation is shown in Figs. \ref{fig_lgwnc}, \ref{g_l}b where the past light cone is constructed from the null characteristics of the light rays coming from all the stars to the observer, and the future gravity null cone is made of the null characteristics of the gravity-field wave equation (\ref{he}) that demarcate the region of the causal interaction of the gravity force of the moving planet with any other particles including photons. 

The deflection angle $\a^i$ is an observable and gauge-invariant quantity measured with respect to the unperturbed direction $k^i$ of incoming photon defined in the asymptotically-flat space-time (at past null infinity). For this reason, the relativistic effects associated with the gravitational deflection of light have direct physical interpretation that is discussed in the rest of the present paper.    

\subsection{Snapshot patterns}\l{sp}
Substituting Eqs. (\ref{8})--(\ref{kz}) to Eqs. (\ref{dm}), (\ref{21}) yields
\ba\l{v}
\vec\a_M&=&\a(1+\cos\chi)\vec n\;,\\\l{z}
\vec\a_D&=&\a\,\f{L}{d}\Bigl[({\vec z}\cdot{\vec n})\vec n-({\vec z}\cdot{\vec m})\vec m\Bigr]\;,\\\l{x}
\vec\a_Q&=&\a J_2\f{R^2}{d^2}\Bigl\{\le[({\vec s}\cdot{\vec n})^2-({\vec s}\cdot{\vec m})^2\r]\vec n-2({\vec s}\cdot{\vec n})({\vec s}\cdot{\vec m})\vec m\Bigr\}\\\nonumber
&&+\;\a\,\f{L^2}{d^2}\Bigl\{\le[({\vec z}\cdot{\vec n})^2-({\vec z}\cdot{\vec m})^2\r]\vec n-2({\vec z}\cdot{\vec n})({\vec z}\cdot{\vec m})\vec m\Bigr\}\;,
\ea
where 
\be\l{qk}
\a=\a_{\rm limb}\f{R}{d}\;,
\ee
is the light-ray deflection angle caused by the planetary mass, and
\be
\l{aer}
\a_{\rm limb}=4\le(1-{\vec k}\cdot{\vec v}_P\r)\f{M}{R}
\ee
is the deflection angle on the planetary limb, $L=|{\vec x}_P(s_1)|$ is the absolute value of the displacement of the planetary center of mass from the origin of the coordinate system, the unit vector ${\vec z}={\vec x}_P(s_1)/L$ points from the origin of the coordinate system toward the center of mass of the planet, and all time-dependent quantities like $\vec x_P$, $\vec z$, and $\vec s$, are computed at the retarded time $s_1=t_1-r_1$ corresponding to the time of observation $t_1$. 

It is possible to prove \cite{desf} that Eqs. (\ref{v})--(\ref{x}) are still applicable in the class of scalar-tensor theories of gravity after a formal replacement: $\a\rightarrow\a_\gamma\equiv (1+\gamma)\a/2$, where $\gamma$ is the PPN parameter characterizing the impact of the scalar field on the light propagation \cite{will-livr}. We have also moved the term depending on the first time derivative of the dipole moment from Eq. (\ref{z}) to Eq. (\ref{aer}). This velocity-dependent term is purely radial and corresponds to the relativistic correction of the planetary mass $M$ caused by the radial Doppler shift. This gravimagnetic correction correlates with the PPN parameter $\gamma$ setting a natural limit on the precision of its measurement from a single-epoch experiment. In general relativity, $\gamma=1$, and the maximal general-relativistic deflection of light on the planetary limbs are respectively: $\a_{\rm limb}\simeq 16280$ $\mu$as for Jupiter, and $\a_{\rm limb}\simeq 5772$ $\mu$as for Saturn \cite{cm,bkk}. Taking into account in Eq. (\ref{aer}) the gravimagnetic correction to the total mass makes these deflections modulated up to a few $\mu$as, depending on the relative orbital velocity of the planet and the observer. 

It is worthwhile to emphasize that the entire theory of the microarcsecond light-ray deflection given in this paper is gauge-independent and Lorentz-invariant, that is all equations are valid in any arbitrary frame moving with respect to the original coordinate system $x^\a=(t,x^i)$ with velocity $\b^i$. Equations (\ref{v})--(\ref{x}) of the light-ray deflection angles remain invariant with the velocity $v^i$ replaced to $v^i-\b^i$. Lorentz-invariance of Eqs. (\ref{v})--(\ref{x}) can be easily understood if one remembers that all quantities entering these equations are referred to the events connected by the null cones shown in Figs. \ref{fig_gwnc}, \ref{fig_lgwnc} that are, of course, invariant with respect to the Lorentz transformations. 

\begin{figure*}
\includegraphics{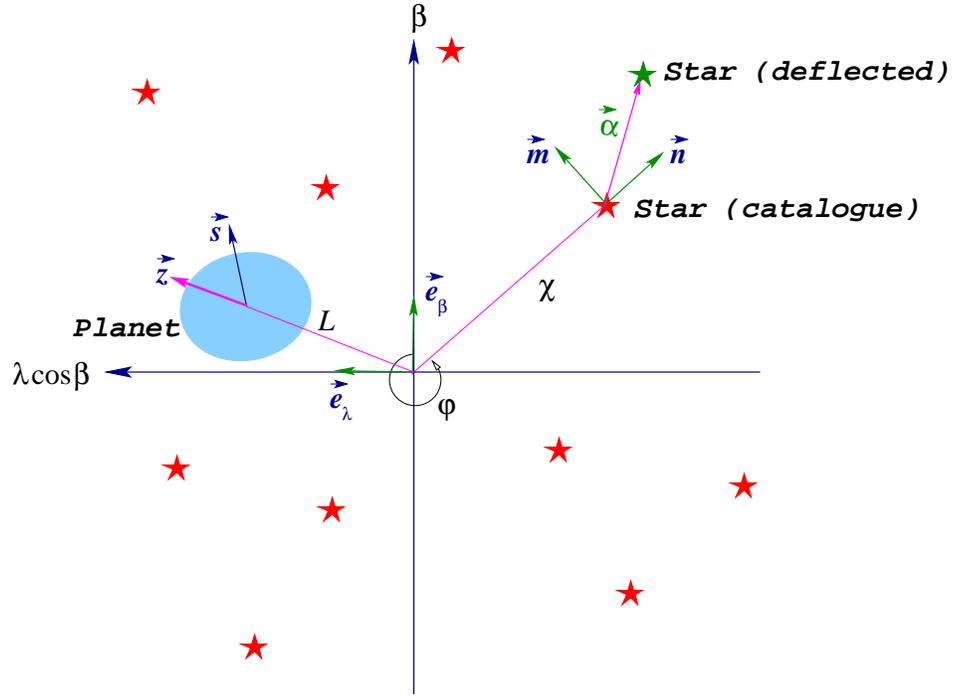}
\caption[Ecliptic coordinates in the plane of the sky]{\label{fig_2}
Local tangential coordinate system in the plane of the sky related to the global ecliptic coordinate system is shown. The unit vectors ${\vec e}_\lambda$ and ${\vec e}_\b$ point in the direction of increasing ecliptic longitude and latitude, respectively. The angular distance of the undeflected position of the star from the origin of the coordinate system is $\chi$ (c.f. Fig. \ref{fig_1}), which is at position angle $\varphi$ from the north direction. The total deflection ${\vec\a}$ has both radial, $\a_n$, and orthoradial, $\a_m$, components: ${\vec\a}=\a_n{\vec n}+\a_m{\vec m}$.}
\end{figure*}

Eqs. (\ref{v})--(\ref{x}) define the deflection patterns caused respectively by the monopolar, dipolar, and quadrupolar components of the planetary gravitational field with respect to the coordinate system $x^\alpha=(t,x^i)$. Because velocity of motion of planets in the Solar system are much smaller than the speed of light/gravity we can assume that during the time of propagation of photon across the Solar system the maximal displacement $x_P^i$ of the center of mass of the planet that is allowed by the ephemeris error or some other violation in the data processing computational approach, is smaller than the impact parameter, $L/d<1$. Let us consider the light-ray deflection patterns for a single epoch of observation, $t_1$, mapping the deflections in the plane of the sky around the planet like it will appear in a photographic snapshot. To vizualize the instantaneous two-dimensional deflection patterns, it is instructive to project Eqs. (\ref{v})--(\ref{x}) on the directions that are parallel and orthogonal to the ecliptic plane of the Solar system as shown in Fig. \ref{fig_2}. The unit vectors $n^i$ and $m^i$ are expanded in such local tangential coordinates into 
\ba\l{22}
{\vec n}&=&{\vec e}_\beta\cos\varphi+{\vec e}_\lambda\sin\varphi\;,\\\l{24}
{\vec m}&=&-{\vec e}_\beta\sin\varphi+{\vec e}_\lambda\cos\varphi\;,
\ea
where $\varphi$ is called the position angle \cite{mar}, the unit vectors ${\vec e}_\lambda$ and ${\vec e}_\beta$ are mutually orthogonal and directed along the increasing ecliptic longitude $\lambda$ and latitude $\beta$ respectively. The unit vectors $\vec z$ and $\vec s$ are expanded in this tangential frame as follows
\ba\l{xx1}
{\vec z}&=&{\vec e}_\beta\cos\psi+{\vec e}_\lambda\sin\psi\;,\\\l{xx2}
{\vec s}&=&{\vec e}_\beta\cos\omega+{\vec e}_\lambda\sin\omega\;,
\ea
where $\psi$ and $\omega$ are the position angles of the unit vectors $\vec z$ and $\vec s$ respectively.

Substituting Eqs. (\ref{22})--(\ref{xx2}) to Eqs. (\ref{v})--(\ref{x}) yields 
\ba\l{vy1}
\vec\a_M&=&\a(1+\cos\chi)\left({\vec e}_\beta\cos\varphi+{\vec e}_\lambda\sin\varphi\r)\;,\\\l{zy1}
\vec\a_D&=&\a\,\f{L}{d}\Bigl[{\vec e}_\beta\cos(\psi-2\varphi)-{\vec e}_\lambda\sin(\psi-2\varphi)\Bigr]\;,\\\l{xy1}
\vec\a_Q&=&\a J_2\f{R^2}{d^2}\Bigl[{\vec e}_\beta\cos(2\psi-3\varphi)-{\vec e}_\lambda\sin(2\psi-3\varphi) \Bigr]\\\nonumber
&&+\;\a\,\f{L^2}{d^2}\Bigl[{\vec e}_\beta\cos(2\omega-3\varphi)-{\vec e}_\lambda\sin(2\omega-3\varphi)\Bigr]\;.
\ea
In this form, the equations clearly show that the directions of deflection angles ${\vec\a}_M$,  ${\vec\a}_D$ and ${\vec\a}_Q$ vary proportional to $\sin\varphi$, $\sin2\varphi$ and $\sin3\varphi$, respectively, as one goes around the planetary center. The amplitudes of $\a_M$, $\a_D$, and $\a_Q$ fall off as $1/\chi$, $1/\chi^2$, and $1/\chi^3$, respectively, where $\chi$ is the impact angle in the plane of the sky between the star and the  coordinate system origin (see Fig. \ref{fig_1}). 
  
\begin{figure*}
\includegraphics{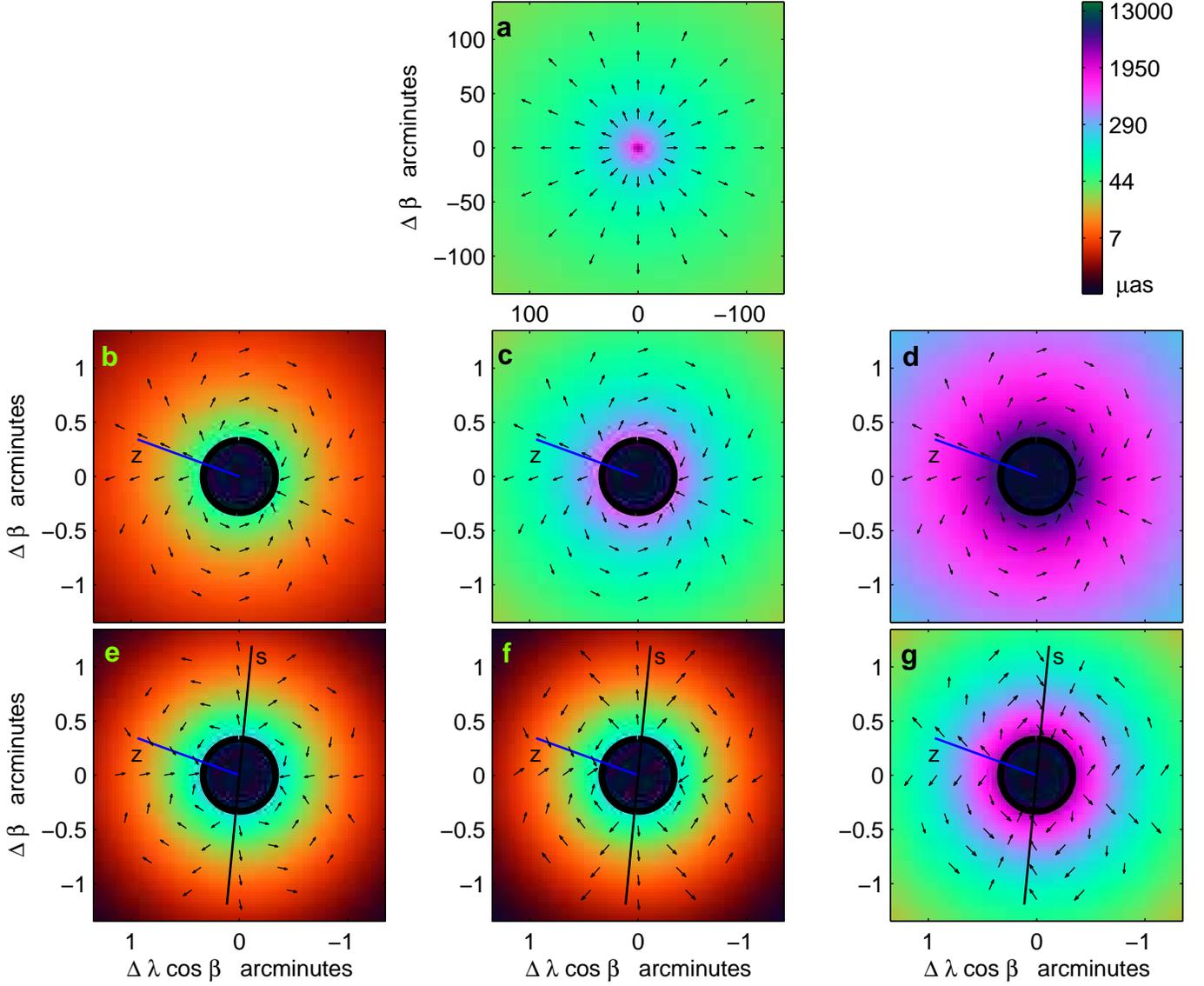}
\caption[Light deflection by Jupiter]{\label{fig_3}
Gravitational deflection of light by the multipolar fields of the Jupiter. The magnitude of deflection is coded by colors in a
logarithmic scale shown at the top. The vectors in each plot indicate only the direction of displacement, not the magnitude.
The monopole deflection pattern is shown in plot (a). The artifactual dipolar deflection patterns caused by a positional
displacement $\vec{x}_P=L\,\vec{z}$ of the planet's center of mass from the origin of the coordinate system used for multipolar expansion of the gravitational field, are depicted in the
central row of plots, for $L_1=350$ (b), $L_2=3500$ (c), and $L_3=35000$ km (d). The bottom row of plots (e--g) shows the combined quadrupolar
deflection patterns generated by the {\it intrinsic} oblateness of the planet and by the coordinate-dependent translational
moment, for the same values of $L$. The direction of rotation axis $\vec s$ is indicated with a thin black line. Note that the quarupolar deflection
at $L_1=350$ km is dominated by that from the {\it intrinsic} moment, whereas for $L_3=35000$ km, on the contrary, the translational quadrupole exceeds the light deflection caused by $J_2$.}\end{figure*}

\begin{figure*}
\includegraphics{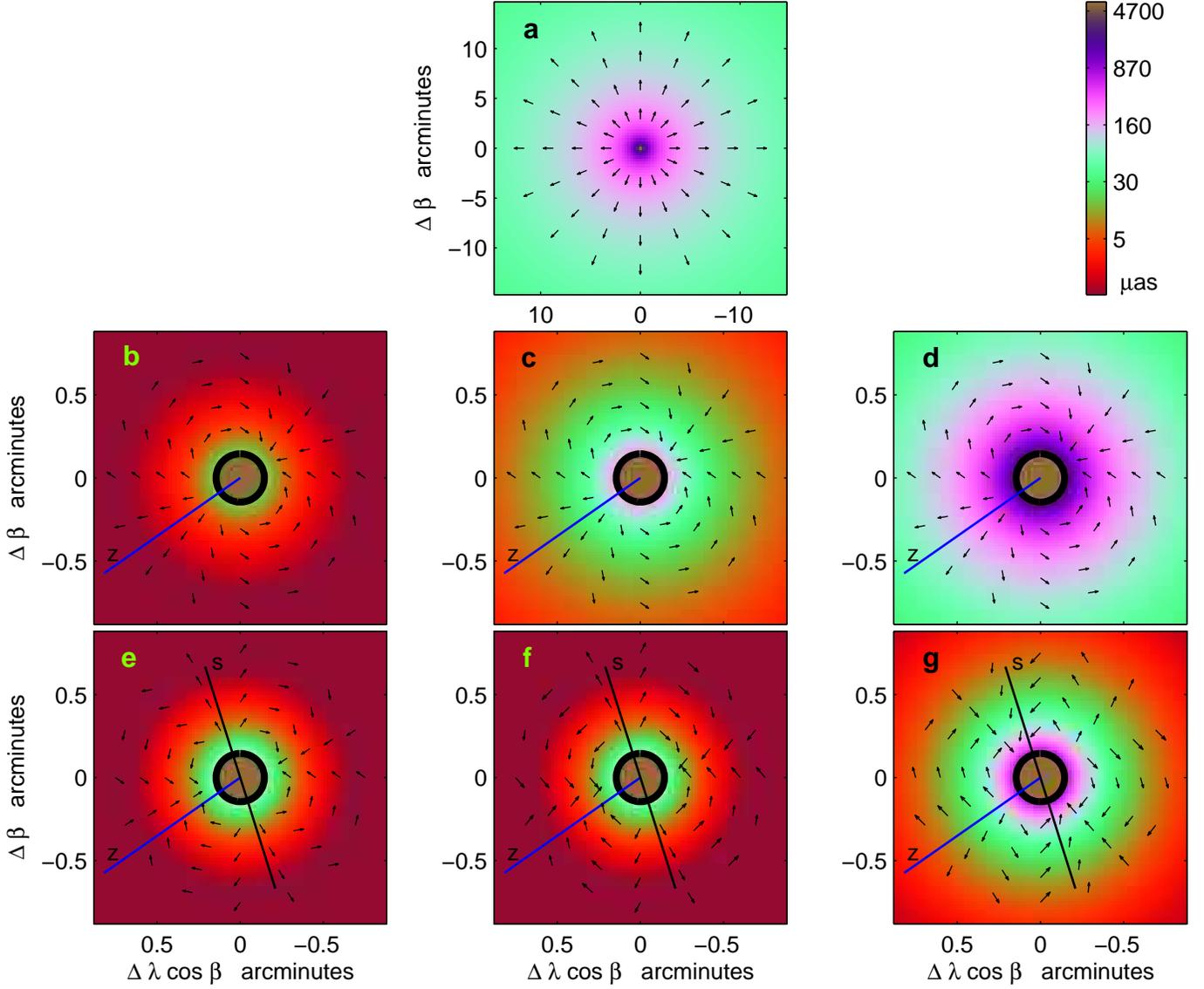}
\caption[Light deflection by Saturn]{\label{fig_4}
Deflection of light by the Saturn caused by its mass (a), dipole (middle raw), and quadrupole (bottom raw) moments. Magnitudes of the deflection are expressed by colors in the logarithmic color grade shown in the upper right corner, and the direction (not the magnitude) of the deflection is shown by small arrows. Vector displacement $\vec L= L \vec z$ of the center of mass of Saturn from the origin of the coordinate system takes values (from left to right) of $L_1=250$ (b), $L_2=2500$ (c), and $L_3=25000$ km (d). The unit vectors $\vec z$ and $\vec s$ indicate the directions of the displacement $L$ and rotational axis of the planet respectively.}
\end{figure*}

Snapshot patterns of deflections ${\vec\a}_M$, ${\vec\a}_D$, and ${\vec\a}_Q$ for a stellar field surrounding Jupiter and Saturn are shown in Figs. \ref{fig_3}, \ref{fig_4} where we use color grades to denote the magnitude of the deflection as a function of the impact angle $\chi$ of the light ray, and arrows to specify the direction of deflection, which depends on the position angle $\varphi$ in accordance with Eqs. (\ref{vy1})--(\ref{xy1}). The patterns are shown for three different magnitudes $L$ of the displacement vector $\vec x_P$: $L_1=350$ km (left), $L_2=3500$ km (middle), $L_3=35000$ km (right) for Jupiter, and $L_1=250$ km (left), $L_2=2500$ km (middle), and $L_3=25000$ km (right) for Saturn. 

The monopolar deflection of light $\vec{\a}_M$ is depicted in subplots (a) of Figs. \ref{fig_3}, \ref{fig_4}. It is purely radial (that is aligned with vector $\vec n$), and does not depend on the magnitude and direction of the displacement vector $\vec x_P$. The dipolar deflection of light, $\vec\a_D$, is directly proportional to the magnitude $L$ of the displacement, and its orientation is determined by vector $\vec x_P$ in the plane of the sky. The quadrupolar deflection of light, $\vec\a_Q$, is quadratically proportional to the magnitude $L$ of the displacement between the origin of the coordinate system and the planet's center of mass as well as on the magnitude of the {\it intrinsic} quadrupole moment of the planet $J_2$. The orientation of the quadrupolar deflection pattern is determined by vector $\vec x_P$ and the rotational axis of the planet. When the displacement $L\ll\sqrt{J_2}R$, the effect of the translational quadrupole $Mx_P^{<i}x_P^{j>}$ on the light deflection is relatively small, and the orientation of the quadrupolar pattern is defined by $\vec s$ as shown in subplots (e) in Figs. \ref{fig_3}, \ref{fig_4}. If $L\simeq\sqrt{J_2}R$, the quadrupolar pattern may be complicated and hard to interpret as the deflection of light by the translational quadrupole $Mx_P^{<i}x_P^{j>}$ is comparable to that caused by $J_2$. If, for any reason, $\sqrt{J_2}R\ll L$, the deflection of light by the {\it intrinsic} quadrupole moment of the planet is swamped by the much stronger deflection caused by the translational quadrupole as shown in subplots (g)  in Figs. \ref{fig_3}, \ref{fig_4}. 

\subsection{Dynamic patterns}
\begin{figure*}
\includegraphics{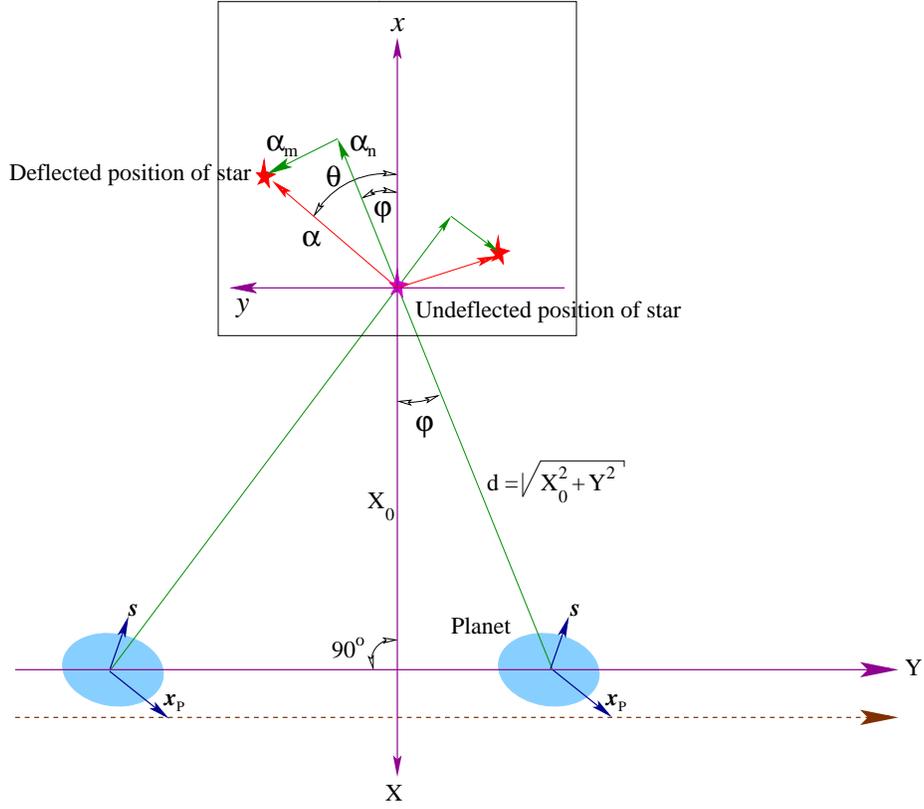}
\caption[Central mapping of light deflection]{\label{map}
Geometry of the central mapping used for computation of the dynamic patterns (curves) of the gravitational deflection of the apparent position of a star by a planet as a function of time. The origins of the two coordinate charts $(x,y)$ and $X,Y$ coincide and placed at the undeflected position of the star. For the sake of graphical convenience, Y axis is displaced along the positive direction of X axis at the constant distance $X_0$. The available planetary ephemeris predicts that planet moves continuously from left to right along $Y$ axis with $X$ fixed as time passes, and two positions of the planet are shown to explain the mapping geometry. The dashed line is parallel to $Y$ axis and shows the trajectory of the real planetary center of mass which may differ from the planetary ephemeris based on the other (independent) set of observations giving rise to the dipolar component of the light deflection. The minimal distance between the planet and the undeflected position of the star is $X=X_0$. Apparent position of the star is deflected from its unperturbed (catalogue) position in both radial and orthoradial directions that are $\a_n$  and $\a_m$ respectively. The total deflection $\vec\a=\a_n\vec n+\a_m\vec m$. The mapping establishes a one-to-one correspondence $F$ between the planet's coordinates $(X,Y)$ and the star's deflected position $(x,y)$, that is $(F:$ $d\rightarrow\a$, $\varphi\rightarrow\theta)$.  }
\end{figure*}

\l{dpc}
Eqs. (\ref{vy1})--(\ref{xy1}) can be used to work out a dynamic visualization of the deflection patterns as a function of time reckoned at the point of observation. Let us assume that a planet is traversing on the celestial sphere below a star along a great circle with a minimal separation angle $\chi_{\rm m}$. As the deflection angles are rather small we can project this motion onto the plane of the sky so that it represents a straight line. We further assume for simplicity that this straight line is parallel and sufficiently close to the ecliptic. This is a good approximation since the orbital inclinations of Jupiter and Saturn are small and do not exceed $2.5^\circ$ \cite{not4}. Eqs. (\ref{vy1})--(\ref{xy1}) can be viewed as a one-to-one mapping of the planetary position specified by coordinates $(X,Y)$ in the plane of the sky, to the locus of the star image shifted by the planetary gravitational field, at coordinates $(x,y)$ (see Fig. \ref{map} and reference \cite{crv} for a more detailed geometry of the mapping). The undeflected position of the star (when the planet is at "infinity") is the center of the mapping which is also the origin of both coordinate frames: $(X,Y)$ and $(x,y)$ (notice that we have shifted Y axis for making better graphical visualization). In other words, the undeflected position of the star is at the point $x=0$, $y=0$ where we also have $X=0$, $Y=0$. The planet moves along a straight line parallel to Y axis at a constant separation $X=X_0$ from the undeflected position of the star. When the planet is at position angle $\varphi$ in the coordinates $(X,Y)$, the deflected position of the star is at angle $\theta$ in the coordinates $(x,y)$. Notice that in general, $\theta\not=\varphi$ because the dipolar and quadrupolar light-ray deflections are not purely radial. Precise correspondence between the two angles is established by the central mapping transformation equations. The angle $\varphi$ runs from $-\pi/2$ to $+\pi/2$ while the angle $\theta$ runs in the most common case from $0$ to $2\pi$ (see Fig. \ref{map} and the text below).

\begin{figure*}
\includegraphics{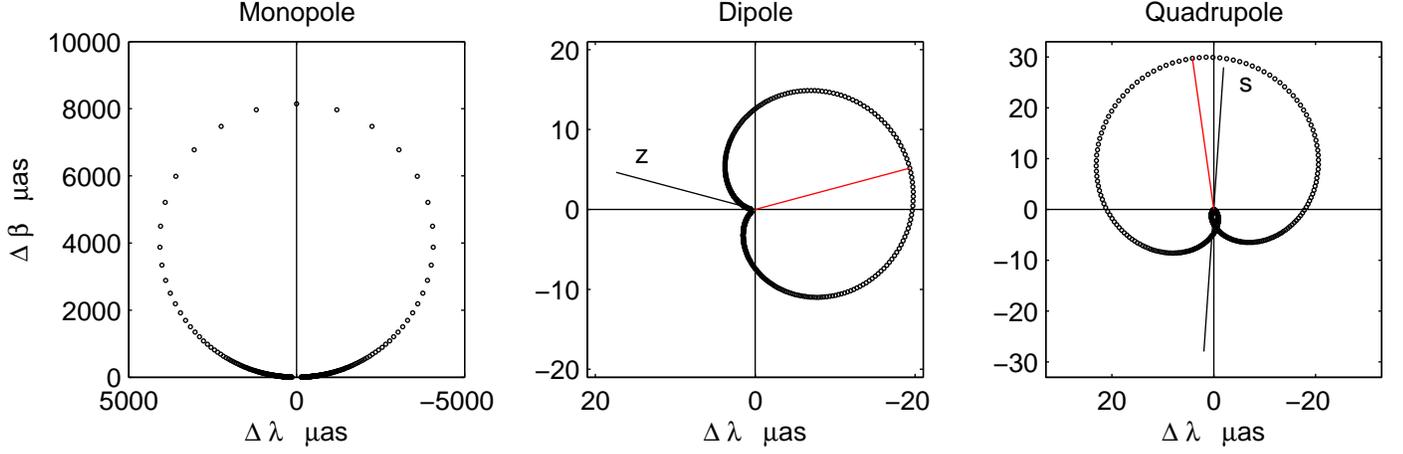}
\caption[Dynamic curves of the light deflection]{\label{fig_5}
Trajectories of apparent positions of a star deflected by the Jupiter. Left, middle and right graphs show respectively the monopolar (circle), dipolar (cardiod), and quadrupolar (the Caley's sextic) light-ray deflection curves. These curves are obtained by central mapping transformation explaind in Fig. \ref{map} and section \ref{dpc}. The planet passes directly below the true position of the star at the coordinate origin, at a constant velocity. The impact parameter $X_0=40$ arcsecond, and the orbital error is $L=250$ km. The time step between individual points is $0.02$ days for the monopole deflection curve, and $0.002$ days for the other two. Dynamic light-ray deflection curves for the Saturn have smaller amplitude but similar shape.}
\end{figure*}

Substituting the deflection angle $\vec\a_M\equiv (x,y)$, $\sin\varphi=Y/d$, $\cos\varphi=X/d$, $d=\sqrt{X^2+Y^2}$ to Eq. (\ref{vy1}), and approximating for simplicity $\cos\chi=1$, we draw the monopolar deflection mapping equations in the following form
\be\l{mcur}
x=2r\;\f{X^2}{X^2+Y^2}\;,\qquad\qquad y=2r\;\f{XY}{X^2+Y^2}\;,
\ee
where $r\equiv 2M/X$, so that $2r$ is the maximal deflection angle reached at the time when the angular distance between the planet and the star is minimal ($\chi_{\rm m}$). The mapping given by Eq. (\ref{mcur}) is purely radial which means that the deflected position of the star and the planet lie on a straight line passing through the center, and $\theta=\varphi$ (see Fig. \ref{map}).  
As the planet moves from $\varphi=-\pi/2$ to $\varphi=+\pi/2$ along the $Y$ axis, the deflected position of the star describes the circle 
\be\l{cir}
(x-r)^2+y^2=r^2\;,
\ee
with radius $r$ and the origin located at the point $(x=r,y=0)$. 
In polar coordinates, $x=\rho\cos\theta$, $y=\rho\sin\theta$, and the correspondence between $\theta$ and $\varphi$ is $\theta=\varphi$, as we have pointed out earlier.  The monopolar deflection curve in polar coordinates is
\be\l{cir1}
\rho=2r\cos\varphi\;,
\ee
which yields the circle shown in the left graph of Fig. \ref{fig_5}, where $x=\Delta\b$, $y=\Delta\lambda$. 

The dynamic curve of the dipolar deflection can be easier represented in coordinates $(x',y')$ that are rotated clockwise with respect to the $(x,y)$ coordinates through angle $\psi$, where $\psi$ is defined by Eq. (\ref{xx1},
\be\l{rcor}
x'=x\cos\psi-y\sin\psi\;,\qquad\qquad y'=x\sin\psi+y\cos\psi\;.
\ee
In the new coordinates, the dipolar mapping transformation is deduced from Eq. (\ref{zy1}) and is given by
\be\l{dcur}
x'=2p\;\f{X^2(X^2-Y^2)}{(X^2+Y^2)^2}\;,\qquad\qquad y'=4p\;\f{X^3Y}{(X^2+Y^2)^2}\;,
\ee
where $p\equiv rL/X$. The dipolar mapping generally contains an orthoradial deflection component as demonstrated in Fig. \ref{map}. As the planet moves from $\varphi=-\pi/2$ to $\varphi=+\pi/2$ along $Y$ axis, the deflected position of the star describes the curve 
\be\l{card}
(x'^2+y'^2-px')^2=p^2(x'^2+y'^2)\;,
\ee 
shown in the middle graph of Fig. \ref{fig_5}. This curve, called cardioid \cite{crv}, is symmetric around the axis $x'$ with a cusp at the origin as shown in the middle graph of Fig. \ref{fig_5}. In polar coordinates, $x'=\rho\cos\theta'$, $y=\rho\sin\theta'$, the correspondence between $\theta'$ and $\varphi$ is $\theta'=2\varphi$, and in these coordinates the cardioid curve takes the form
\be\l{card1}
\rho=p\le(1+\cos 2\varphi\r)\;.
\ee 
Its orientation depends on the angle $\psi$, and the magnitude on the ratio $L/X$. Notice that the vector ${\vec x}_P$ was assumed to be constant. In reality, the planet may have a residual transverse velocity with respect to its ephemeris trajectory. In this case, the dipolar deflection curve will slightly deviate from the cardioid.  

The dynamic trajectory of the quadrupolar deflection becomes simple in coordinates $(\hat x,\hat y)$  rotated clockwise with respect to the $(x,y)$ coordinates through angle $2\psi$,
\be\l{rco}
\hat x =x\cos 2\psi-y\sin 2\psi\;,\qquad\qquad \hat y=x\sin 2\psi+y\cos 2\psi\;.
\ee
In these new coordinates, the quadrupolar mapping transformation may now be inferred from Eq. (\ref{xy1})
\be\l{dcu}
\hat x =4q\;\f{X^4(X^2-3Y^2)}{(X^2+Y^2)^3}\;,\qquad\qquad \hat y=4q\;\f{X^3Y(3X^2-Y^2)}{(X^2+Y^2)^3}\;,
\ee
where the parameter $q\equiv q_L= (r/2)(L/X)^2$ for the deflection component caused by the translational quadrupole $Mx^{<i}_Px^{j>}_P$, and $q\equiv q_J=(r/2)J_2(R/X)^2$ for the term generated by the {\it intrinsic} quadrupolar oblateness $J_2$ of the planet. The quadrupolar mapping contains an orthoradial deflection besides the radial component. As the planet moves from $\varphi=-\pi/2$ to $\varphi=+\pi/2$ along $Y$ axis, the deflected position of the star describes the curve 
\be\l{sec}
4({\hat x}^2+{\hat y}^2-q\hat x)^3=27q^2({\hat x}^2+{\hat y}^2)^2\;,
\ee 
which is shown in the right graph of Fig. \ref{fig_5}. This curve, called the Caley's sextic \cite{crv}, is symmetric around the axis ${\hat x}$ and has a small secondary loop near the origin. The polar equation of the sextic is gained by  substitution of the polar coordinates $\hat x=\rho\cos\hat \theta$, $\hat y=\rho\sin\hat \theta$ in Eq. \ref{sec}, and solving the cubic equation with respect to $\rho$. The quadrupolar mapping correspondence between the angles $\hat\theta$ and $\varphi$ is $\hat\theta=3\varphi$, and the deflection caused by the translational quadrupole
\be\l{sec1}
\rho=q_L\le(\cos 3\varphi+3\cos\varphi\r)\;.
\ee 
The sextic curve for the deflection caused by the oblatness $J_2$, takes the same form after replacing $q_L\rightarrow q_J$ and $\psi\rightarrow\omega$. The quadrupolar deflection curve has a secondary loop passing through the origin of coordinate axes as shown in Fig. \ref{fig_5}. The cross-over point of the secondary loop is 1/8 of the maximum deflection $\rho=4q_L$ in the mid-point of the primary loop. This means that the magnitude of the secondary loop is about few microarcseconds for Jupiter that is withing the range of experimental measurement by SIM and SKA. 

\begin{figure*}
\includegraphics{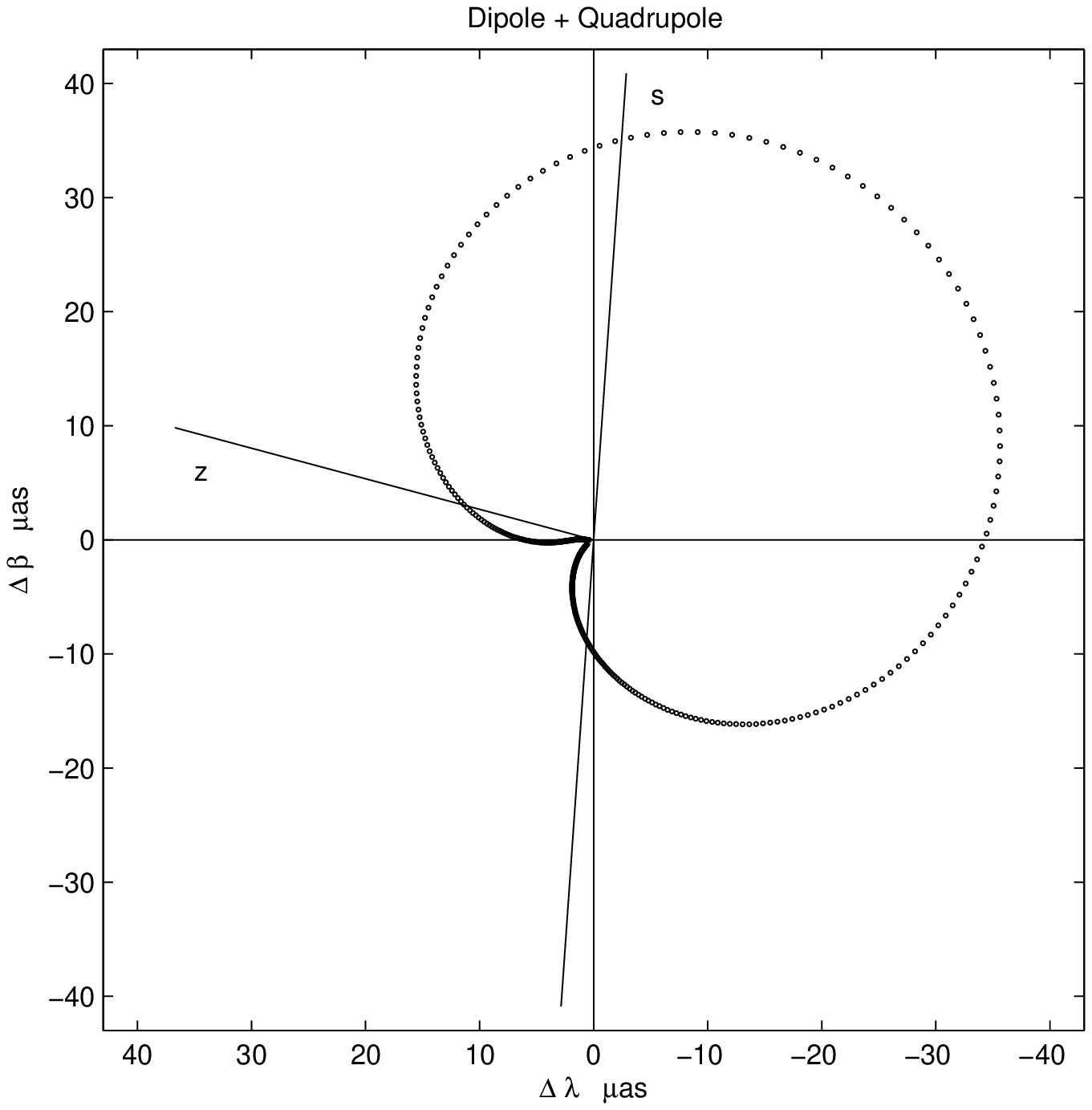}
\caption[Combined dipolar and quadrupolar deflection curves]{\label{fig_6}
Combined dipolar and quadrupolar deflection of light by the Jupiter for the same parameters and configuration as in 
Fig. \ref{fig_5}. Details of the quadrupolar deflection (the sextic curve) are hidden by the dipolar deflection (the cardioid). Separation of the two deflection
terms is a non-trivial experimental problem, and generally will require multiple observations of a number of sources around
the planet.}
\end{figure*}

Fig. \ref{fig_6} shows the combined deflection trajectory from the dipole cardioid and the quadrupole sextic presented in
Fig. \ref{fig_5}. The resulting curve looks similar to a cardioid of a somewhat larger amplitude, and is inclined at some intermediate angle with respect to the cardioid and sextic symmetry axes. Obviously, this curve swamps the quadrupolar deflection and can be confused with
a pure dipolar one, because the vector $\vec{x}_P$ is not known beforehand. This demonstrates the crucial importance of
accurate determination of the spurious dipole component in any experiment designed to measure the {\it intrinsic} gravitational
quadrupole $J_2$. Since the ephemeris error is expected to be a smooth function of time, microarcsecond observations of
multiple stars at different position angles to the direction of the dipole axis should help to separate these two contributions to the total
deflection pattern.
  
\section{Experimental Tests of Relativity and Reference Frames}\l{ore}
Crosta and Mignard \cite{cm} assumed that as light propagates toward the observer the gravitating planet's center of mass is at rest at the origin of the chosen intermediate coordinate system in the sky. Our formalism stems from a more general dynamical interpretation, because it takes into account the effect of planet's motion on the deflection angle in a rigorous relativistic paradigm. The planetary motion appears in the light-ray deflection measurements in two ways:
\begin{itemize}
\item explicitly -- in the form of velocity-dependent term, ${\vec v}_P$, and acceleration, ${\vec a}_P$, of the planet with respect to the coordinate system of our choice;
\item implicitly -- in the form of the retarded time, $s_1=t_1-r_1$, in the coordinates, ${\vec x}_P={\vec x}_P(s_1)$ of the planet as well as in its velocity and acceleration.
\end{itemize}
General relativity may hypothetically be violated in the higher-order terms of approximation, which will become detectable in the near future by astrometric facilities operating at the 1 $\mu$as level of precision. This presumable violation of general relativity can be parametrized and tested in light deflection experiments. Verification of general relativity and the quest for beyond-GR phenomena include the search for scalar fields, observation of gravimagnetic and retardation-of-gravity effects associated with the motion of planet, detection and measurement of the dipolar and quadrupolar anisotropy, and improvement of fundamental reference frames and planetary ephemerides in the Solar system. In what follows, we discuss these experimental opportunites in more detail by analysing different modes of the gravitational light-ray deflection. 

\subsection{The monopolar deflection}

The monopolar deflection of light is given by Eqs. (\ref{v}), (\ref{qk}), (\ref{aer}). It is standard experimental practice to introduce the PPN papameter $\gamma$ leading to the replacement \cite{will-livr}
\be\l{repl}
\a\rightarrow\a_\gamma\equiv \f{1+\gamma}{2}\a\;
\ee 
in Eq. (\ref{v}) as well as Eqs. (\ref{z}), (\ref{x}).
The parameter $\gamma$ characterizes deviation of gravity from a pure geometry and is associated with the presence of hypothetical scalar fields remaining from the epoch of the Big Bang \cite{dnor}. This parameter is usually measured in the solar light-ray deflection experiments with a precision approaching to the level of $10^{-4}\div 10^{-5}$ \cite{kspv,bit}. 

Orbital motion of a planet generates $h_{0i}$ component of the gravimagnetic field which depends on the time derivative of the dipole moment as appears in Eq. (\ref{5}). This gravimagnetic component leads to the {\it explicit} dependence of the monopolar light-ray deflection on the radial component of the orbital velocity ${\vec v}_P$ of the planet, effectively modulating the planetary mass as shown in Eq. (\ref{aer}). 
The gravimagnetic term was first derived in \cite{ks} and analytically confirmed in \cite{kl-aa,frit,wuk}. It can reach about 1.6 $\mu$as for a star observed at the limb of the Jupiter. 

The gravimagnetic term is the result of the asymmetric exchange of energy between the photon and the gravitational field of the moving body. It is well-known \cite{br} that in case of a static gravitational field the overall angle of light deflection does not depend on the energy (frequency) of the photon. The photon gains energy as it approaches the gravitating body, and loses it as it moves away. If the body is at rest with respect to the chosen coordinate system, the gain and the loss of energy cancel out, but this balance is violated if the body is moving radially. Appearance of the Doppler modulation of the deflection of light can be also understood if one remembers that the deflection angle is associated with spatial components of the null vector of photon. Lorentz transformation from a static frame of the planet to the moving frame of observer transforms and mixes all four components of the wave vector and its time component affects the spatial components in the moving frame. This is clearly shown in the paper by Klioner \cite{kl-aa}. 

The gravimagnetic deflection of light can be measured in a single-epoch observation only if (1) mass of the planet is known sufficiently well; (2) the parameter $\gamma$ is excluded since it correlates linearly with the gravimagnetic term. On the other hand, if we can conduct observations in different epoches, the gravimagnetic term in Eq. (\ref{aer}) will periodically modualte the radial deflection because of the relative motion of the planet with respect to observer. Hence, it can be measured independently of the parameter $\gamma$. In case of the Jupiter the amplitude of this gravimagnetic modulation reaches 3.2 $\mu$as and has a main period of one year due to the orbital motion of the Earth.

\subsection{The dipolar deflection}\l{dart}

The dipolar deflection of light is given by Eqs. (\ref{z}), (\ref{qk}) where the 
coordinates of the planetary center of mass ${\vec x}_P={\vec x}_P(s_1)$ depend on the retarded instant of time $s_1=t_1-r_1$, which is a solution of the retarded null-cone equation (\ref{rt}). These retarded coordinates of the planet define the dipole moment $\vec I(s_1)=M{\vec x}_P(s_1)$ in the multipolar expansion of the gravitational field of the planet. In general relativity, the law of conservation of linear momentum allows us to eliminate the dipole moment by placing the center of mass of the planet at the origin of the coordinate system. 

If ${\vec x}_P(s_1)=0$, all coordinate-dependent effects in the gravitational deflection of light vanish. It means that the planet deflects light from the retarded position defined by the retarded solution of the gravity wave equation (\ref{he}) where $t=t_1$ is the time of observation and $r=r_1$ is the distance between the observer and the planet as shown in Fig. \ref{fig_lgwnc}. This retardation is due to the fact that light and gravity interact on the hypersurface which is the intersection of two null cones - the past light cone of observer and the future gravity cone of the moving planet. This important prediction of general relativity \cite{kapjl} was tested and confirmed in 2002 \cite{fk,kf,kf-pla} with precision of $\sim 20\%$. SIM, Gaia, and SKA can improve this measurement by, at least, a factor of 10.
In a practical experimental setup, the retardation of gravity effect can be parametrized by introducing a parameter $\epsilon$ to the retarded time, $s\rightarrow s_\epsilon=t-\epsilon\, r$ \cite{not5} in the dipolar light-ray deflection angle (\ref{z}). If $\epsilon=0$, the gravitational field of the planet would deflect photons instantaneosly. The general relativity prediction is that $\epsilon=1$, which corresponds to the case of gravity propagating with the same speed as light. This consideration and comparision with the calculations of Crosta and Mignard \cite{cm} make it evident that the position of the Jupiter's center of mass, that is not specified in the paper \cite{cm}, must be taken at the retarded instant of time $s_1=t_1-r_1$ with respect to observer. 

For arbitrary $\epsilon\not=0$ (that is when the speed of gravity and light differ) light is deflected by the planet from its orbital position taken at the retarded time $s_\epsilon$. It leads to a non-vanishing displacement vector $\vec L$ 
\be\l{eps}
\vec L\equiv\vec x_P(s_1)-\vec x_P(s_\epsilon)=(\epsilon-1)\vec v_P r_1-\f12(\epsilon-1)^2\vec a_P r^2_1+O\le[(\epsilon-1)^3\r]\;,
\ee
where $\vec v_P$ and $\vec a_P$ are velocity and acceleration of the planet respectively.
If gravity operates on a  null cone that is different from the light cone, it brings about the dipolar anisotropy 
\ba\l{qs}
\vec \a_D&=&(\epsilon-1)\a\,\f{r_1}{d}\Bigl[({\vec v}_P\cdot{\vec n})\vec n-({\vec v}_P\cdot{\vec m})\vec m\Bigr]\\\nonumber
&-&\f12(\epsilon-1)^2\a\,\f{r^2_1}{d}\Bigl[({\vec a}_P\cdot{\vec n})n^i-({\vec a}_P\cdot{\vec m})m^i\Bigr]\;,
\ea
which can be measured to evaluate the retardation of gravity parameter $\epsilon$ \cite{kf,kf-pla,k-ijmp,k-pla}. For a star observed on the limb of Jupiter ($d=R$ where $R$ is the radius of the planet) the velocity-dependent term in Eq. (\ref{qs}) is $\sim\a_{\rm limb}(v/c)(r_1/R)\simeq 8140$ $\mu$as, and the acceleration-dependent term is $\sim\a_{\rm limb}(v^2/c^2)(r_1/2R)\simeq 0.2$ $\mu$as. The reader should notice that the dipolar deflection (\ref{qs}) depends on the transversal component of the planetary velocity only. Since the velocity of the planet is almost lying in the ecliptic plane, it makes the position angle $\psi=-\pi/2$ of the unit vector $\vec z=\vec L/L$ in Eq. (\ref{xx1}) so that the symmetry axis of the cardioid in Fig. \ref{fig_5} gets aligned with $Y$ axis. This also allows us to physically interpret the dipolar deflection associated with the retardation of gravity as caused by the gravimagnetic field of the Jupiter originating from its translational orbital motion \cite{arak,ser,scb,ij1,k-pla}. 

A VLBI experiment was conducted in 2002 \cite{fk} to search for the residual dipolar deflection of light from the QSO J0842+1835 by the moving gravitational field of the Jupiter, caused by the presumable difference between the speed of gravity and light parametrized by $\epsilon$. This experiment did not show any deviation from the general relativistic model of light deflection by the moving planet in excess of 20\%.

Assuming that general relativity is fully valid, measurements of the dipolar anisotropy of light deflection can be used to determine the position of the center of mass of the planet on its orbit more accurately than all other currently available astrometric techniques (see section \ref{sss} for numerical estimates).

\subsection{The quadrupolar deflection}
It is evident from Eq. (\ref{x}) that in order to measure the quadrupolar deflection of light caused by the planetary oblatness $J_2$, the following condition on the dispacement vector $\vec x_P$ must be satisfied
\be\l{25}
|\vec x_P|\leq|\vec\sigma_L|<\sqrt{J_2}R\;,
\ee 
where $\sigma_L$ is the measurement error of $\vec x_P$. If condition (\ref{25}) is fulfilled, all terms in the second line of Eq. (\ref{x}) depending on the ratio $(L/d)^2$ can be abolished. In fact, $\sigma_L\not=0$ because the planetary ephemeris is known with a limited accuracy. Hence, one should consider the vector $\vec x_P$ as a fitting parameter that is to be determined from light-ray deflection observation by successive iterations. JPL ephemerides provides the initial value of $\vec x_P$ for such iterations, which can be improved if the measurement accuracy on the deflected positions is high enough (see next section). For the Jupiter $\sqrt{J_2}R \simeq 8500 km$, so that in order to measure the {\it intrinsic} quadrupole moment $J_2$ of the planet its orbital position must be known to $\sigma_L<8500 km$. The currently available JPL ephemerides for major Solar system bodies, determined from direct optical/radio observations and spacecraft tracking, are believed to be more accurate. However, astrometry of gravitational light deflection provides another independent method of refining the dynamic model of the Solar system. A non-zero value of $L$ also emerges if  the time of observation $t_1$ instead of the retarded time $s_1=t_1-r_1$ is used  in the light-ray deflection equations whenever the instantaneous position of the planet is computed, which would be the case of $\epsilon=0$ in Eq. (\ref{eps}). Then, $L\approx 35000 km$ for the Jupiter, and condition (\ref{25}) is violated, thus, making the {\it intrinsic} quadrupolar deflection of light unmeasurable. 

After the displacement $\vec x_P$ is determined from the estimated dipole anisotropy, one can shift the origin of the coordinate system in the sky to the center-of-mass of the light-ray deflecting planet to supress the spurious dipolar term as much as possible. If general relativity is a correct theory of gravity, this will reduce both the dipolar and quadrupolar anisotropies caused by the translational displacement $\vec L$ of the planetary center of mass. On the other hand, if general relativity is violated one may not be able to remove the dipole anisotropy because the structure of the light-ray deflection equations will be different from those shown in Eqs. (\ref{v})--(\ref{x}). This would make it more difficult to measure $J_2$ in the models of the light deflection based on an alternative theory of gravity.

\section{Observational Capabilities}\l{ocap} 
Three astrometric missions capable of angular measurements at about 1 $\mu$as level accuracy are under development for the
relatively near future. Two of them, the Space Interferometry Mission (SIM) \cite{sim} and Gaia \cite{gaia}, are space-based
optical facilities, while the Square Kilometer Array (SKA) \cite{ska} is a ground-based radio telescope array. We discuss in more detail the SIM and SKA since the gravitational applications of Gaia has been described in \cite{cm}.
\subsection{SIM optical interferometry}\l{sss}

Perhaps, the SIM,
which is a Michelson-type interferometer with articulating siderostat mirrors, holds the best prospects
for precision tests of general relativity in the Solar system through gravitational bending effects. In these experimental-gravity applications
the adavantages of the SIM facility are as follows:
\begin{enumerate}
\item unlike Gaia, SIM is a pointing mission which can be aimed at an interesting object at the most
appropriate time. Planetary near-limb grazing passages of sufficiently bright stars will be necessary to confidently
measure the monopole light-ray deflections from all the major planets, and the dipole and quadrupole moments from the Jupiter and, presumably, the Saturn. Such conjunction events are fairly rare but can be accurately predicted. 
\item SIM is primarily designed for astrometric detection of terrestrial-mass planets. In the differential regime of 
operation the interferometer is expected to achieve the unprecedented accuracy of 1 $\mu$as in a single observation on
stars separated in the sky within $\sim$2 degrees. A single measurement accuracy of Gaia is more than an order of magnitude poorer even for the brightest stars.
\item the baseline of SIM can be rotated through 90 degrees for a dedicated observation, providing nearly
simultaneous two-dimensional observations of a set of stars while the light-ray deflecting planet is still near them. Two-dimentional
observations on a given set of stars are crucial for unambigious disentangling the dipole and quadrupole deflection patterns,
which may have similar magnitudes in the vicinity of a planetary limb.
\item SIM will self-calibrate its $15^\circ$-wide field of regard to rms 20-40 pm in Zernike polynomial
coefficients, which corresponds to the range of 0.4 - 0.9 $\mu$as subtended on the sky. This dramatically reduces
the correlated and/or systematic errors and allows us to extend precision of
differential measurements at 1 $\mu$as to wider angles in the central part of the field of regard, if needed.
\item owing to the small field stop in the optical design, SIM can observe stars and quasars as close as several arcseconds from the planetary limb, that is for impact parameter of the light ray comparable with the planetary radius.
\end{enumerate}

Assuming that four stars are observed with SIM at roughly the same impact angular distance $\chi$ from Jupiter's
center of mass, the signal-to-noise ratio (S/N) in measurement of the dipole deflection angle $\alpha_D$ from a single one-dimensional
observational visit is
\be\l{x1}
\le[\f{S}{N}\r]_{\a_D}=0.0254f \le[\f{1\;\mu{\rm as}}{\sigma_0}\r]\le[\f{1'}{\chi}\r]^2\biggl[\f{z}{1\;{\rm km}}\biggr]\;,
\ee
where $\sigma_0$ is the single measurement differential error, and $(f\leq 2)$ is a geometric factor that depends on the actual position of the stars and the angle between the baseline of the SIM interferometer and the direction of vector $\vec x_P$ characterizing the dipolar light-deflection pattern (see Fig. \ref{fig_1}). The measurement error $\sigma_L$ on the magnitude of $L=|\vec x_P|$ for this configuration is
\be\l{x2}
\sigma_L = 39.37f^{-1} \le[\f{\sigma_0}{1\; \mu{\rm as}}\r]\biggl[\f{\chi}{1'}\biggr]^2\quad{\rm km}.
\ee
Thus, the best achievable accuracy in determination of $L$ (the error in the position of the planetary center of mass) with SIM is roughly 20 km in the sky plane, when
$\chi = 1'$ and $\sigma_0= 1$ $\mu$as. This is more than 10 times better than the currently available positional ephemeris
for the Jupiter \cite{st,pt1,pt2}. However, it will be hardly possible to achieve this
accuracy in a single measurement, because of the difficulty to find simultaneously four stars within $1'$ of the planetary limb. Multiple
observations of different star configurations should be planned. 

Assuming that the radial distance between the observer and the Jupiter is 5 AU, the amount of the gravitational quadrupole deflection of light from the planet is roughly 
\be\l{x3}
\alpha_Q = 8.9 \le[\f{1'}{\chi}\r]^3\quad\mu{\rm as}\;,
\ee
as follows from the $J_2$-dependent term in Eq. (\ref{x}).
It will be accurately measured with SIM if fairly dense fields of background stars can be found along the path of the
planet in the sky. For the previously considered idealized configuration of four equally-distant stars, two one-dimesional
observations with orthogonal baselines will be required to disentangle the dipolar and quadrupolar components of the light-ray deflection pattern. For a pair of mutually orthogonal observations, the signal-to-noise ratio in measurement of the quadrupole deflection angle $\a_Q$ is
\be\l{x4}
\le[\f{S}{N}\r]_{\a_Q}=8.9f \le[\f{1\; \mu{\rm as}}{\sigma_0}\r]\le[\f{1'}{\chi}\r]^3\;,
\ee
where again $(f\leq 2)$ is a geometric factor. Interestingly, the quadrupolar deflection from the Jupiter is
comparable to the expected dipolar deflection at $\chi = 1'$, but since $\a_Q$ falls off faster than $\a_D$ as the angular distance $\chi$ of a star from the planet increases, multiple observations at wider angles from the planetray limb may be 
advantageous to separate and
to determine the dipolar component of the light-ray deflection. Because of the complicated pattern of the quadrupolar component, a combined
reduction of all observed events will be required, after the dipolar correction will have been determined and taken into account. 

\subsection{SKA radio interferometry}

Extensive discussion of various fascinating science drivers and of the evolving technical possibilities has led to a concept for the Square Kilometre Array (SKA) and a set of design goals \cite{ska}. The SKA will be an interferometric array of individual antenna stations, synthesizing an aperture with a diameter of up to several 1000 kilometers. A number of configurations will distribute the 1 million square meters of collecting area. These include 150 stations each with the collecting area of a 90 m telescope and 30 stations each with the collecting area equivalent to a 200 meters diameter telescope. The sensitivity and versatility of SKA can provide $\sim$ 1 $\mu$as astrometric precision and high quality milliarcsec-resolution images
by simultaneously detecting calibrator sources near the target source if an appreciable component of SKA is contained in elements which are more than 1000 km from the core SKA \cite{freid}. 

Measurement of the light bending by a moving planet with microarcsecond accuracy requires a continuous phase-referencing observation of the target and the calibrating radio sources \cite{fk,fkp}. The main limitation of the accuracy is the tropospheric refraction which affects radio observations.  The
large-scale tropospheric refraction can be estimated by observing many
radio sources over the sky in a short period of time.  At
present the determination of the global troposphere properties can
only be estimated in about one hour, and smaller angular-scale
variations can not be determined in most cases. However, the SKA, by using observations in ten subarrays, on strong radio sources around the sky, will determine the tropospheric properties on time-scales which may be as short at five minutes. 

Quasars as astrometric calibrators have one peculiar property: they are variable.  The radio emission from the quasar comes from the base of the jet which is formed by an accretion disk around a massive black hole. The massive outflows
and shocks in the jet change the intensity and the structure
of the radio emission.  Hence, the position of the quasar reference point (a radio
core) is variable by about 0.05 mas in most quasars \cite{sud03}.  Thus,
the calibrators used to determine the SKA astrometric precision to
better than $10~\mu$sec have jitter which is somewhat larger.  In order to reach the intended angular precision, the change in position of the calibrators must be
determined.  With knowledge of the evolution of radio cores and the
monitoring of a basic set of primary calibrators the
calibrator grid can be determined, at least to $10~\mu$as or better \cite{freid}.  The use of
many calibrators in the field of view would also diminish the net effect of position jitter.

In addition to various special and general relativistic effects in the time
of propagation of electromagnetic waves from the quasar to the SKA-VLBI antenna
network, we must account for the effects produced by the planetary
magnetosphere \cite{ronek,fk-mag}. To be more specific, let us concentrate on the case of the Jupiter. 
Consideration of the magnetospheric effects for the Saturn can be done similarly.

Measurements obtained during the occultations of Galileo by
Jupiter indicate \cite{jupmag} that near the surface of Jupiter the
electron plasma density reaches the peak
intensity $N_0 = 1.0\times 10^{10}$ m$^{-3}$. We shall assume that the jovian
magnetosphere is spherical, though in reality the magnetosphere has a dipole
structure and our model may underestimate the
plasma content along the polar directions. We postulate that a radial drop-off of the plasma density $N(r)$ is proportional
to $1/r^{2+A}$ where
$r$ is the distance from the center of Jupiter, and $A$ is a number. The guess is that
$A\geq 0$, and we will assume that $A=0$ for the worst possible case. Hence,
radial dependence of the electron plasma density is taken as
$N(r)=N_0(R_J/r)^{2+A}$, where $R_J=7.1\times 10^7$ m is the mean radius of
Jupiter.

The plasma produces a delay $\Delta T$ in the time of propagation of radio
signal which is proportional to
the column plasma density in the line of sight given by integral
\cite{yak}
\begin{eqnarray}
\label{p}
N_l&=&\int^{r_0}_d \frac{N(r)\,dr}{ r^{A+1}\sqrt{r^2-d^2}}+\int^{r_1}_d
\frac{N(r)\,dr}{ r^{A+1}\sqrt{r^2-d^2}}\;,
\end{eqnarray}
where $r_0$ and $r_1$ are radial distances of quasar and radio antenna from
Jupiter respectively, and $d=|{\bm \xi}|$ is the impact parameter of the
light ray from the quasar to Jupiter. In the experiment
under discussion the impact parameter is much less than both $r_0$ and $r_1$.
Hence,
\begin{eqnarray}
\label{pp}
N_l(\mbox{m}^{-2})&=&N_0R_J\left(\frac{R_J}{
d}\right)^{A+1}\frac{\sqrt{\pi}\Gamma\left(\frac{A+1}{2}\right)}{
\Gamma\left(1+\frac{A}{ 2}\right) }\;,
\end{eqnarray}
where $\Gamma(z)$ is the Euler gamma-function.
The plasma time delay (in seconds)
\begin{eqnarray}
\label{ppp}
\Delta T(\mbox{s})&=&40.4\,c^{-1}\nu^{-2} N_l\;,
\end{eqnarray}
where $c$ is the speed of light in vacuum measured in m/sec, $\nu$ is the
frequency of electromagnetic signal measured in Hz.

It is worthwhile noting that, in fact, the SKA-VLBI array measures difference in
path length between the radio telescopes. Hence, one has to differentiate
$N_l$ in expression (\ref{ppp}) with respect to the impact parameter $d$ and
project the result on the plane of the sky. This gives a magnetospheric VLBI
time delay of
\begin{eqnarray}
\Delta_{\rm mag}\mbox{(ps)}&=&6.3\times 10^{-7}(A+1)\frac{N_l}{
d}\left(\frac{\nu_0}{ \nu}\right)^2\frac{{\bm n}\cdot{\bm b}}{ c}\;,
\end{eqnarray}
normalized to the frequency $\nu_0=8.0$ GHz. The deflection of light $\delta\alpha_{\rm mag.}$ caused by the magnetosphere is evaluated as $\delta\alpha_{\rm mag.}\simeq c\Delta_{\rm mag}/b$ rad.

For example, substituting $d=5R_J$ and
taking the SKA-VLBI baseline $b=6000$ km we find
\begin{eqnarray}
\label{qqq}
\delta\alpha_{\rm mag.}&\simeq&163\left({\nu_0/\nu}\right)^2\; \mbox{$\mu$as}\qquad\quad(A=0)\;,
\\
\delta\alpha_{\rm mag.}&\simeq&24\left({\nu_0/\nu}\right)^2\; \mbox{$\mu$as}\qquad\quad(A=1)\;,
\\
\delta\alpha_{\rm mag.}&\simeq&14\left({\nu_0/\nu}\right)^2\; \mbox{$\mu$as}\qquad\quad(A=2)\;.
\end{eqnarray}
This represents the pure bending from Jupiter's magnetosphere at a single radio frequency $\nu_0=8$ GHz, which should be compared with the magnitude of the gravitational deflection of light by various multipoles at the light-ray closest approach. SKA higher frequency is expected to be 35 GHz \cite{ska1}. At this frequency the magnetospheric deflection is smaller as compared to 8 GHz by a factor of 20.  

In any case, the magnetospheric deflection estimate reveals that a single frequency observation of the light deflection will be affected by the magnetosphere at the level exceeding 1 $\mu$as. This assumes that we should observe at two widely spaced frequencies to determine and eliminate the magnetospheric effects. However, to keep sensitivity at higher frequencies two polarizations and rather wide bandwidth must be used. The noise due to turbulence in the magnetosphere (and the Earth ionosphere) may also be a limit. However, this rapidly fluctuation model is
fairly pessimistic and unlikely, and would probably average out to the
steady state model. The most optimum method to deal with the possible
jovian magnetosphere component is not yet known and should be a matter of special study in radio astronomical community.

\newpage
\newpage
\newpage
\newpage
\newpage
\newpage
\newpage
\newpage
\newpage

\begin{thebibliography}{}
\bibitem{ander} Quadrupolar fields of the Solar system planets and their natural satellites can be independently measured with the Doppler tracking technique as well \cite{titan}. The quadrupolar field causes the gravitational red shift of the frequency of a microwave beam sent to and transponded back from an interplanetary spacecraft to be a complicated function of the quadrupole deformation.
\bibitem{not} Notice that $t_*$ changes its value from one inertial frame to another in accordance with the law of the Lorentz transformation. Eq. (\ref{oku}) gives $t_*$ in a static, non-moving frame.
\bibitem{not1} Real observer (SIM, Gaia, VLBI station, etc.) moves with respect to the coordinate system $x^\a$ with velocity $\b^i$. Relativistic effects of this motion are obtained by making use of the Lorentz transformation technique \cite{kk,klio2,kl-aa}.
\bibitem{nnq} Finite distance effects in the light-ray deflection angle may be important in the solar gravitational bending of light for several nearby stars (see equation 208 and accompanying discussion given in our paper \cite{ks}).
\bibitem{mnot} Papers \cite{asad,sam} follow the line of reasoning adopted in \cite{will-livr,will-b} and does not distinguish the past null cone of light and the future null cone of gravity, thus, confusing the causal property of gravity (speed of gravity) with that for light.
\bibitem{not4} In real astrometric experiment the inclicnation of the planetary orbits to the ecliptic must be taken into account.
\bibitem{not5} Rigorous mathematical parametrization of the retardation of gravity effect is given in \cite{kop-cqg,ij1}. It changes the field equation (\ref{he}) to 
\be\l{he1}
\le(-\epsilon^2\f{\p^2}{\p t^2}+{\vec\nabla}^2\r)h_{\a\b}=0\;,
\ee
so that both the gravitational field solution (\ref{4})--(\ref{6}) and the retarded position of the planet in the light-ray deflection angle $\a$ become dependent on the retarded time $s_\epsilon=t-\epsilon r$.
\end{thebibliography}
\end{document}